\documentclass[aip,graphicx]{revtex4-1}

\usepackage{amssymb}
\usepackage{amsmath}
\usepackage{siunitx}
\usepackage{graphicx}
\usepackage{dcolumn}
\usepackage{bm}
\usepackage{soul}
\usepackage{changes}
\usepackage{array}
\usepackage{multirow}
\usepackage{xcolor}

\usepackage[utf8]{inputenc}
\usepackage[T1]{fontenc}
\usepackage{mathptmx}
\usepackage{etoolbox}

\begin{document}


\title{Design optimization of silicon nitride nanomechanical resonators for thermal infrared detectors: a guide through key figures of merit} 



\author{Paolo Martini}
\affiliation{Institute of Sensor and Actuator Systems, TU Wien, Gusshaustrasse 27-29, Vienna, 1040, Austria}
\author{Kostas Kanellopulos}
\affiliation{Institute of Sensor and Actuator Systems, TU Wien, Gusshaustrasse 27-29, Vienna, 1040, Austria}
\author{Silvan Schmid}
 \email{silvan.schmid@tuwien.ac.at}
\affiliation{Institute of Sensor and Actuator Systems, TU Wien, Gusshaustrasse 27-29, Vienna, 1040, Austria}


\date{\today}

\begin{abstract}
Thermomechanical infrared (IR) detectors have emerged as promising alternatives to traditional photon and thermoelectric sensors, offering broadband sensitivity and low noise without the need for cryogenic cooling. Despite recent advances, the field still lacks a unified framework to guide the design of these nanomechanical systems. This work addresses that gap by providing a comprehensive design guide for IR thermal detectors based on silicon nitride drumhead and trampolines. Leveraging a validated analytical model, we systematically explore how geometry, tensile stress, and optical properties influence key performance metrics such as thermal time constant, noise-equivalent power, and specific detectivity. The analysis encompasses both bare silicon nitride and structures with broadband absorber layers, revealing how different parameter regimes affect the trade-off between sensitivity and response speed. Rather than focusing on a single device architecture, this study maps out a broad design space, enabling performance prediction and optimisation for a variety of application requirements. As such, it serves not only as a reference for benchmarking existing devices but also as a practical tool for engineering next-generation IR sensors that can operate close to the fundamental detection limit. This work is intended as a foundational resource for researchers and designers aiming to tailor IR detectors to specific use cases.
\end{abstract}


\maketitle 

\section*{Legend}

\section{Introduction}
\label{intro}


Infrared (IR) detection technologies fall into two categories: photon detectors and thermal detectors. While photon detectors such as HgCdTe photodetectors offer high sensitivity, their operation is fundamentally limited by the photon energy, making them suitable primarily for narrow band applications in the short-wave and lower mid-infrared (MIR) spectral regions \cite{kruse1997uncooled}. Moreover, they typically require cryogenic cooling, increasing system complexity and cost \cite{rogalski2019infrared}. As an alternative, thermal IR detectors, including bolometers, pyroelectric detectors, and thermopiles, operate well at room temperature and provide sensitivity across a wide spectral range. However, these thermoelectric detectors are fundamentally constrained by electrical noise sources, such as Johnson noise and $1/f$ noise, which ultimately limit their sensitivity.
Thermomechanical detectors, on the other hand, offer a compelling route to circumvent these limitations by exploiting the effect of absorbed thermal energy on the mechanical properties of a resonator, such as changes in stiffness, which are then read out through its mechanical response rather than electrical one, thereby reducing electrical noise. Since their first introduction \cite{vig1996uncooled}, uncooled thermomechanical detectors, particularly nanomechanical systems, have emerged as promising alternatives by transducing temperature changes into measurable mechanical resonance shifts \cite{zhang2013nanomechanical, hui2013high, vicarelli2022micromechanical, piller2022thermal, li2023terahertz, zhang2024high, das2023thermodynamically, zhang2025enhanced}.

Silicon nitride (SiN) is currently one of the materials of choice for nanomechanical sensing, owing to its exceptional mechanical stability and thermomechanical properties, such as a high thermal expansion coefficient. In addition, in the specific context of IR detection, it exhibits favourable optical properties. Its absorption spectrum at room temperature features a pronounced peak coinciding with the characteristic emission of an ideal black body, effectively functioning as a metamaterial absorber. At the same time, it possesses a broad transparency window spanning from the visible to the mid-infrared, making it well suited for IR spectroscopy applications, and a high refractive index that can be tuned through stoichiometry control~\cite{beliaev2022optical}. A SiN-based detector can be easily rendered broadband by depositing an absorber, sacrificing only a small fraction of its sensitivity while substantially extending its detectable wavelength range. Recent advances in nanofabrication, including the development of low-stress, uniform SiN films and precise sub-micron patterning techniques, have enabled the realisation of ultra-thin, low-loss drumheads and more complex resonator architectures, extending the detection limits of these devices. 

Among the available thermomechanical resonator architectures, drumhead and trampoline geometries are particularly advantageous, as their increased active areas enhance coupling to long-wavelength radiation while preserving key performance metrics \cite{kanellopulos2025comparative}. State-of-the-art SiN drumheads have achieved noise-equivalent power (NEP) on the order of $\mathrm{pW/\sqrt{Hz}}$ in both narrowband \cite{zhang2024high} and broadband \cite{martini2025uncooled} applications, approaching the fundamental room-temperature detectivity limit of $D^*_{lim}=$\SI{1.8e10}{\cm\sqrt\Hz\per\watt} for a resonator thermally coupled to the environment from only one side \cite{datskos2003detectors, kruse2004can, skidmore2003superconducting}. Recently, Zhang \textit{et al} \cite{zhang2024high} reported an $NEP\approx\SI{36}{\pico\watt\per\sqrt\hertz}$ and a detectivity o f $D^*\approx\SI{3.4e9}{\centi\m\sqrt\hertz\per\watt}$ using a metamaterial drumhead optimised for the range \SIrange[]{0.3}{5}{\tera\hertz}. The structure has a thermal response time of $\approx\SI{200}{\milli\s}$. Using a free-space impedance matched (FSIM) metal absorber for broadband IR light detection \cite{martini2025uncooled} has lead to similar results of $NEP=\SI{27}{\pico\watt\per\sqrt\hertz}$ and $D^*=\SI{3.8e9}{\centi\m\sqrt\hertz\per\watt}$. The metal layer helped reduce the thermal constant to $\tau_{th}=\SI{14}{\milli\s}$. Trampoline resonators have shown excellent performance, too. Piller \textit{et al}. \cite{piller2022thermal} reached a NEP as low as \SI{7}{\pico\watt\per\sqrt\hertz} using SiN trampoline resonators featuring a similar absorber as in \cite{martini2025uncooled}, with a response time of only \SI{4}{\ms}. More recently, Das \textit{et al.} \cite{das2023thermodynamically} proposed a multi-material trampoline structure incorporating a perforated absorber optimised for long-wave infrared detection, reaching a specific detectivity of $D^* =\ $\SI{3.8e9}{\cm\sqrt\Hz\per\watt} with \SI{7.4}{\ms} thermal response. Another work \cite{vicarelli2022micromechanical} made use of SiN trampoline resonators with a Cu and Au layer as an absorber and reported an $NEP\approx\SI{100}{\pico\watt\per\sqrt\hertz}$ in sub-THz (\SI{0.14}{\tera\hertz}) detection, and a $\tau_{th} = \SI{4}{\milli\s}$. In a very recent work \cite{zhang2025enhanced}, bare SiN trampolines achieved a record $D^* = \SI{6.4e9}{\cm\sqrt\Hz\per\watt}$ with $\tau_{th} = \SI{88}{\milli\s}$. Operating at the noise level set by the temperature-fluctuation limit, the authors were able to maintain a detectivity only a factor of three lower ($D^* = \SI{2.5e9}{\cm\sqrt\Hz\per\watt}$) while achieving a faster response time ($\tau = \SI{3}{\milli\s}$) than that dictated by the resonator’s thermal time constant.

Despite these advancements, a unified framework for optimising detector parameters remains absent in the field. This work addresses this gap by presenting a comprehensive analysis based on a validated analytical model \cite{kanellopulos2025comparative, schmid2023fundamentals, Kanellopulos2025dissertation}, and by systematically exploring the resonator parameter space. Focusing on SiN drumheads and trampolines, we investigate how variations in geometry, intrinsic stress, and absorbivity influence the thermal and mechanical performance of the devices, The resulting trade-offs are evaluated using key figures of merit: thermal time constant ($\tau_{th}$), noise equivalent power (NEP), and specific detectivity ($D^*$). 

Our model reconciles recent experimental results \cite{piller2022thermal, kanellopulos2025comparative, martini2025uncooled}. The framework enables performance prediction across design geometries, providing practical guidelines for developing application-specific IR detectors. In particular, we examine whether, in light of the current state-of-the-art, existing devices are already operating near their fundamental limits or whether opportunities for notable performance gains remain. The framework presented here enables this assessment quantitatively, highlighting potential pathways to approach the fundamental photon noise limit.

\section{Figures of Merit}
\label{sec:FOM}

\begin{figure}
    \centering
    \includegraphics[width=\linewidth]{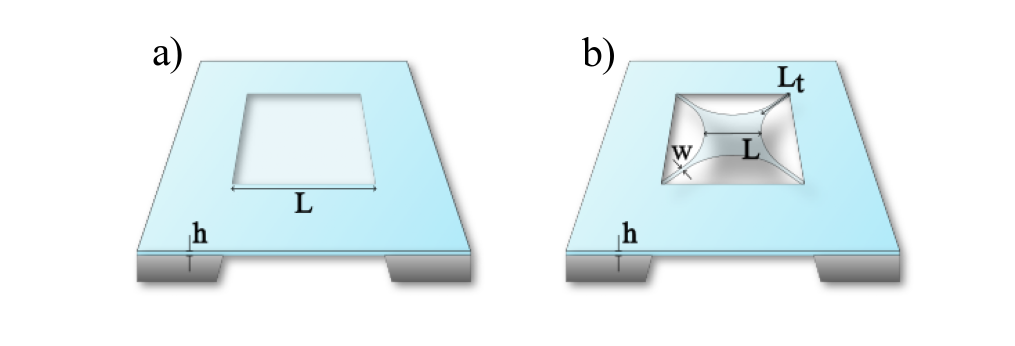}
    \caption{Schematic representation (not in scale) of the two designs under investigation in this work. a) square drumhead with the geometric variables studied highlighted: $h$ is the thickness of the SiN layer, hence of the structure, and the lateral size $L$. In addition to these two, for a b) trampoline we study also the influence of the tethers' width, $w$, and length, $L_t$.}
    \label{fig:schematic}
\end{figure}

Before reviewing the metrics used to evaluate the detectors’ performance, we introduce the two designs considered in this work. Regardless of the specific fabrication process employed, we assume the detector chip to consist of a silicon (Si) substrate several hundred \SI{}{\micro\m} thick. This substrate serves both as a mechanical support for handling the resonator and as a thermal reservoir where heat generated by absorbed radiation can be effectively dissipated. A layer of silicon nitride (SiN), deposited on top of the silicon substrate with low-pressure chemical vapour deposition (LPCVD), forms the material of the detecting element. Figure \ref{fig:schematic} presents a schematic overview of the two geometries. In Figure \ref{fig:schematic}a, a square drumhead is illustrated, defined by two primary geometrical parameters: its thickness $h$ and lateral dimension $L$. Compared to this basic configuration, the trampoline design introduces additional geometrical complexity, as shown in Figure \ref{fig:schematic}b. Here, the variable $L$ denotes the side length of the central suspended area, which serves as the active detection region, analogous to the drumhead's side length. This central area is connected to the supporting frame via four tethers, each characterised by a length $L_T$ and a width $w$. These four parameters ($h$, $L$, $L_T$, and $w$) fully define the geometries of the drumhead and trampoline structures, whose impact on detector performance is systematically explored in Section \ref{sec:geometry}.\newline

\subsection{Thermal time constant, $\tau_{th}$}
\label{sec:FOM_tau}
The first figure of merit is the thermal time constant, $\tau_{th}$ [$\mathrm{s}$]. This metric is crucial in many applications, particularly when fast detector response is required, e.g.: in time-resolved spectroscopy applications, where scanning frequencies can reach the order of \SI{2}{\kilo\hertz}. While in most cases this remains the fundamental limit of the detection chain, a recent work \cite{zhang2025enhanced} showed that this limitation can be broken when the resonator is operated at its fundamental temperature fluctuation noise limit. 

For a lumped element system, $\tau_{th}$ is defined as the ratio between the heat capacity $C$ [$\mathrm{J/K}$] of the resonator and its thermal conductance $G$ [$\mathrm{W/K}$] \cite{kruse1997uncooled}:
\begin{equation}
\tau_{th} = \frac{C}{G}.
\label{eq:tau}
\end{equation}

For a drumhead made of a material with mass density $\rho$ and specific heat capacity $c_p$, in the context of the mean temperature framework (MTF), the heat capacity is given by \cite{kanellopulos2025comparative}:
\begin{equation}
C = \rho c_p h L^2.
\label{eq:C_drumhead}
\end{equation}
Likewise, the heat capacity of a trampoline structure is \cite{kanellopulos2025comparative}:
\begin{equation}
C =\rho c_p h (L^2 + 4 w L_t).
\label{eq:C_trmapoline}
\end{equation}
By comparing (\ref{eq:C_drumhead}) and (\ref{eq:C_trmapoline}), it is evident that for the same active detection area $L^2$, trampolines exhibit a larger heat capacity due to the additional volume contributed by the tethers.

In this study, we consider detectors operating in vacuum conditions, where heat transfer via convection can be neglected. In this regime, the thermal conductance is the sum of the conductive and radiative contributions only: $G = G_{cond} + G_{rad}$. For a drumhead, this takes the form \cite{kruse1997uncooled}:
\begin{equation}
G = 4\pi h \kappa + 4 L^2 \epsilon \sigma_{SB} T^3_0,
\label{eq:G_drumhead}
\end{equation}
where $\kappa$ is the thermal conductivity of the material, $\epsilon$ is the emissivity, $\sigma_{SB}$ is the Stefan–Boltzmann constant, and $T_0$ is the temperature of the environment including the detector. The first term represents heat conduction through the drumhead, while the second accounts for radiative losses from the surface.\newline
In a trampoline structure, the conductive heat transfer term arises from the series combination of the thermal resistances of a membrane ($R^{m}$) and that of four strings ($R^{s}$) of length $L_T$ and width $w$:
\begin{equation}
    R^{m} + R^{s} = \frac{1}{G^{m}_{cond}}+\frac{1}{G^{s}_{cond}} = \frac{1}{G_{cond}},
    \label{eq:G_cond_tramp}
\end{equation}
where $G^{m}_{cond} = 4 \pi h \kappa$ and $G^{s}_{cond} = \frac{8wh}{L_T}\kappa$ represent the terms for thermal conductance through conduction of a membrane and four strings, respectively  \cite{schmid2023fundamentals}.
The radiative contribution, on the other hand, accounts for the total emitting surface of both the membrane and the tethers. The overall thermal conductance of the trampoline can thus be expressed as:
\begin{equation}
G = \frac{8 \pi h  w}{2w + \pi L_T}\kappa + 4(8 w L_T + 2L^2)\epsilon\sigma_{SB}T^3_0.
\label{eq:G_trampoline}
\end{equation}
The contribution of the tethers to both conduction and radiation is evident from the presence of $w$ and $L_T$ in both terms of the sum.

\subsection{Power responsivity, $R_P$}
\label{sec:FOM_Rp}
When a light source with power $P_0$ impinges on the detector surface, a fraction of that power, denoted $P$, is absorbed by the material and converted into heat:
\begin{equation}
P = \alpha P_0,
\label{eq:P}
\end{equation}
where $\alpha$ is the absorptance, which depends on both the material properties and the wavelength of the incident light. The absorbed power raises the temperature of the detector, inducing thermal expansion of the material. This phenomenon, known as induced stress change, underpins the working principle of thermomechanical IR detectors. As the material expands, the mechanical stiffness of the structure changes, which in turn shifts the resonance frequency $\omega_0$ of the detector.

The second figure of merit considered in this work is the power responsivity, $R_P$, with unit [$\mathrm{1/W}$], defined as the relative change in resonance frequency per unit of absorbed power \cite{schmid2023fundamentals}:
\begin{equation}
R_P(\omega) = \frac{\partial \omega_0}{\partial P} \frac{1}{\omega_0} H_{th}(\omega),
\label{eq:Rp}
\end{equation}
where $H_{th}(\omega)$ is a low-pass filter transfer function that accounts for the finite thermal response time of the resonator \cite{schmid2023fundamentals}:
\begin{equation}
H_{th}(\omega) = \sqrt{\frac{1}{1+(\omega \tau_{th})^2}}.
\label{eq:Hth}
\end{equation}

Since thermomechanical detectors operate by converting absorbed power into heat, effectively functioning as temperature sensors, (\ref{eq:Rp}) can be recast in the form \cite{kanellopulos2025comparative, schmid2023fundamentals}: 
\begin{equation}
R_P(\omega) = \frac{\partial \omega_0}{\partial T} \frac{1}{\omega_0} \frac{\partial T}{\partial P} H_{th}(\omega) = \frac{R_T}{G}H_{th}(\omega),
\label{eq:Rp2}
\end{equation}
where $T$ is the temperature, and $R_T$ is the temperature responsivity. For square drumheads, such as those considered in this study, it takes the form \cite{schmid2023fundamentals}:
\begin{equation}
R_{T} = -\frac{\alpha_{th}}{2(1-\nu)}\frac{E}{\sigma},
\label{eq:Rt_drumhead}
\end{equation}
where $\alpha_{th}$ is the thermal expansion coefficient, $\nu$ the Poisson's ratio, $E$ is the Young's modulus, and $\sigma$ the tensile prestress in the material. The negative sign in (\ref{eq:Rt_drumhead}) reflects the fact that thermal expansion causes a softening of the structure, resulting in a decrease in resonance frequency with increasing temperature.

In the case of trampolines, the temperature responsivity becomes \cite{schmid2023fundamentals}:
\begin{equation}
R_{T} = - \frac{\alpha_{th}}{2}\frac{E}{\sigma}.
\label{eq:Rt_trampoline}
\end{equation}
Here, the factor $(1-\nu)$, which accounts for the in-plane expansion in two directions, is absent due to the uniaxial strain of the tethers.

The ratio $E/\sigma$, common to both cases, is a characteristic feature of tensile mechanical resonators and represents an enhancement factor of the temperature responsivity. Finally, while $R_T$ depends solely on material properties, the detector geometry plays a key role in determining the power responsivity $R_P$, as evidenced by the presence of the thermal conductance $G$ in (\ref{eq:Rp2}).

\subsection{Noise equivalent power (NEP)}
\label{sec:FOM_NEP}
To properly assess the performance of a detector, the noise level of the measurement system must also be taken into account. The next figure of merit is the noise equivalent power (NEP), defined as the signal power required to achieve a unitary signal-to-noise ratio within a one $\mathrm{Hz}$ bandwidth. It has units of [$\mathrm{W/\sqrt{Hz}}$], and for a nanomechanical resonator used as an IR detector, it can be expressed as \cite{datskos2003detectors, schmid2023fundamentals}:
\begin{equation}
    NEP = \frac{\sqrt{S_y(\omega)}}{R_P(\omega)}.
    \label{eq:NEP}
\end{equation}
Here, $S_y(\omega)$ is the fractional frequency noise power spectral density, which quantifies the system’s noise by evaluating the frequency stability of the resonator in thermal equilibrium. From (\ref{eq:NEP}), it is clear that achieving high sensitivity requires minimizing frequency noise to resolve the smallest possible shifts in resonance frequency. 

In this work, we consider three primary sources of frequency noise: thermomechanical noise $S_{y_{thm}}(\omega)$, detection noise $S_{y_d}(\omega)$, and temperature fluctuation noise $S_{y_{th}}(\omega)$. It is also worth mentioning that, among many other possible noise sources, one particularly relevant stems from photothermal back action caused by the relative intensity noise of a readout laser. In optomechanical sensing applications involving an IR absorber, this noise can significantly limit sensitivity \cite{martini2023towards}. To mitigate this effect, the absorber design can be optimized, as demonstrated in \cite{martini2025uncooled}.

Assuming the detection noise is white, its contribution can be related to the thermomechanical noise peak magnitude as follows \cite{schmid2023fundamentals}:
\begin{equation}
    S_{z_d}(\omega) = \mathcal{K}^2 S_{z_{thm}}(\omega_0) = \mathcal{K}^2\frac{4k_BTQ}{m_{eff}\omega^3_0},
    \label{eq:Sdet}
\end{equation}
with the effective mass of the resonator $m_{eff}$, the quality factor $Q$, the resonance angular frequency $\omega_0$, and the Boltzmann constant $k_B$. The coefficient $\mathcal{K}$ represents the relative sensitivity of the transduction system: if it can resolve the thermomechanical noise peak, then $\mathcal{K} < 1$. Thermomechanical and detection noise are typically summed into what is called the additive noise. Using (\ref{eq:Sdet}), the corresponding additive frequency noise PSD takes the form \cite{bevsic2023schemes, demir2021understanding}:
\begin{equation}
    S_{y_\theta}(\omega)= \frac{2}{\omega_0^2}\frac{S_{z_{thm}}(\omega_0)}{z^2_0} \left[\lvert H_{\theta_{thm}}({i}\omega)\lvert^2 + \mathcal{K}^2 \lvert H_{\theta_d}(\mathrm{i}\omega)\lvert^2 \right].
    \label{eq:Sy}
\end{equation}
With $z_0$ the displacement amplitude of the resonator. The two transfer functions $H_{\theta_{thm}}({i}\omega)$ and $H_{\theta_d}(\mathrm{i}\omega)$ are filter functions. They are specific to the measurement loop implemented to read out the detector’s mechanical motions. In the case of a self-sustaining oscillator (SSO) detection scheme, they are \cite{bevsic2023schemes}:
\begin{equation}
\begin{split}
    H_{\theta_{\mathrm{thm}}}(\mathrm{i}\omega) &= \frac{1}{\tau_r} H_L(i\omega)\\
    H_{\theta_{\mathrm{d}}}(\mathrm{i}\omega) &= \frac{1}{\tau_r}\frac{1}{H_r(i\omega)}H_L(i\omega). 
    \label{eq:oscillator-Hs}
\end{split}
\end{equation}
Here, $H_r(\omega)$ is the low-pass transfer function of the resonator, with the mechanical time constant $\tau_r=2Q/\omega_0$, and $H_{L}(\omega)$ is the low-pass transfer function of the oscillator circuit.  It has been shown that the transfer functions for other oscillator schemes, such as a phase-locked loops, are essentially equal to (\ref{eq:oscillator-Hs}). A more detailed discussion about these terms, as well as their derivation for other detection schemes, can be found in \cite{bevsic2023schemes}. 

Equation (\ref{eq:Sy}) shows that noise can be reduced, and the NEP improved, by increasing the vibrational amplitude $z_0$. Ideally, the resonator should be operated at the onset of nonlinearity by pushing the actuation to the critical amplitude $z_{0_c}$ \cite{lu2020nonlinear}:
\begin{equation}
    z_{0_c} = \sqrt{\frac{8}{3\sqrt{3}}}\frac{1}{\sqrt{Q}}\sqrt{\frac{m_{eff}\omega^2_0}{\alpha_{Duff}}},
    \label{eq:z_critical}
\end{equation}
with the Duffing parameter $\alpha_{Duff}$. Notably, if $z_0 = z_{0_c}$ and $\mathcal{K} \ll 1$, then $S_{y_{thm}}(0)$ becomes independent of the quality factor $Q$. This implies that increasing the quality factor does not benefit this specific type of noise. Recently, \textit{Manzaneque et al.} \cite{manzaneque2023resolution} reported on an improvement of frequency stability in mechanical resonators when driven in the nonlinear regime ($z_0 > z_{0_c}$), albeit only for short integration times, which is a regime that typically is difficult to access with slow thermoelastic detectors. 

The effective mass $m_{eff}$, which appears in both (\ref{eq:Sdet}) and (\ref{eq:z_critical}), varies according to the geometry of the device. For the fundamental mode in drumheads, it is given by \cite{schmid2023fundamentals}:
\begin{equation}
    m_{eff} = \frac{1}{4}m_0 = \frac{1}{4}\rho h L^2,
    \label{eq:meff_mbrn}
\end{equation}
while for trampolines, still in the fundamental mode, it becomes \cite{kanellopulos2025comparative}:
\begin{equation}
    m_{eff} = \rho h \left(L^2 + \frac{4wL_t}{2}\right).
    \label{eq:meff_trmp}
\end{equation}

Similarly, the Duffing coefficient in (\ref{eq:z_critical}) takes different forms. For a square drumhead resonator, it is \cite{catalini2021modeling}:
\begin{equation}
    \alpha_{Duff} = \frac{3\pi^4}{64}(n^4+j^4)\frac{Eh}{L^2}\frac{1}{2(1-\nu^2)(1-\nu)},
    \label{eq:duff_mbrn}
\end{equation}
with $n$ and $j$ the modal numbers of the vibrational mode. 
For trampolines, the expression derived by \cite{Kanellopulos2025dissertation} is:
\begin{equation}
    \alpha_{Duff} = \frac{n^4\pi^4}{8}\frac{Ewh}{(2L_t)^3}.
    \label{eq:duff_trmp}
\end{equation}
Since the latter equation has not yet been validated before, we performed measurements using SiN trampoline resonators and verified its accuracy. The results are provided in the Supplementary Material (SM). 

The ultimate sensitivity limit for a nanomechanical thermal detector is set by temperature fluctuation noise $S_{y_{th}}(\omega)$ \cite{schmid2023fundamentals, rogalski2002infrared}. This noise arises from stochastic heat exchange between the detector and the environment and generates white frequency noise. For lumped-element thermal detectors it is given by \cite{vig1996uncooled, zhang2025enhanced, kanellopulos2025comparative}:
\begin{equation}
    S_{y_{th}}(\omega) = \frac{4k_BT^2R_T^2}{G}\left|H_{th}(\omega)\right|^2\left|H_{L}(\omega)\right|^2.
    \label{eq:Sth}
\end{equation}
Here, $G$ is the effective thermal conductance of the resonator. It has been shown that $G$ for drumheads and trampolines is well approximated by (\ref{eq:G_drumhead}) and (\ref{eq:G_trampoline}), respectively \cite{kanellopulos2025comparative}.

\subsection{Specific detectivity ($D^*$)}
\label{sec:FOM_D*}
The sensitivity (\ref{eq:NEP}) of nanomechanical membrane and trampoline resonators generally improves with increasing lateral size ($L$) as long as conductive heat transfer dominates. For larger $L$, radiative heat transfer becomes significant. The extra thermal heat loss channel leads to a deterioration in sensitivity. In the radiative heat transfer regime, when the frequency stability is limited by thermomechanical noise (\ref{eq:Sy}), the sensitivity follows the scaling law
\begin{equation}
   NEP\propto \sqrt{L^3}.
\end{equation}
In the ultimate regime, where frequency stability is determined by temperature fluctuation noise (\ref{eq:Sth}), the scaling law changes to
\begin{equation}
   NEP\propto L,
\end{equation}
as derived below.
This latter dependence matches the well-known scaling for photodetectors, whose sensitivity scales with the square root of the detection area, $\sqrt{A}$ \cite{jones1953performance}. A commonly used figure of merit in such cases is the specific detectivity, $D^*$, with units [$\mathrm{cm\sqrt{Hz}/W}$], which normalizes sensitivity for the detector size. Hence, it is particularly useful for comparing sensors with different detection areas. It is defined as \cite{datskos2003detectors, nudelman1962detectivity}:
\begin{equation}
    D^* = \frac{\sqrt{A}}{NEP}.
    \label{eq:D*}
\end{equation}
It is important to note that, in the case of trampoline resonators, the detection area does not include the tethers. Therefore, for this design, $A=L^2$ just as for drumheads. 

According to Kirchhoff’s law, for a resonator in thermodynamic equilibrium with its surroundings, the emissivity $\epsilon$ equals the absorptance $\alpha$. 

Considering the case where temperature fluctuation noise dominates, the equation (\ref{eq:NEP}) can be expressed in terms of $\epsilon$ \cite{schmid2023fundamentals}: 
\begin{equation}
    NEP = \frac{\sqrt {S_{y_{th}}(\omega)}}{\epsilon R_{P}(\omega)} = \frac{\sqrt{4k_BT^2G}}{\epsilon}\left|H_{L}(\omega)\right|^2.
    \label{eq:NEP_rad}
\end{equation}
It is noteworthy that the thermal low-pass transfer function $H_{th}(\omega)$, which normally shapes both the frequency noise PSD (\ref{eq:Sth}) and the responsivity (\ref{eq:Rp}), no longer appears in (\ref{eq:NEP_rad}). As a consequence, the thermal response time ceases to play a role, and the measurement speed is essentially limited only by the time constant of the oscillator circuit $H_{L}(\omega)$, provided the resonator operates in the temperature fluctuation noise regime. This effect has recently been exploited to enhance measurement speed in nanomechanical trampoline resonators \cite{zhang2025enhanced}.

Assuming further that the resonator is well thermally isolated so that heat transfer occurs predominantly via radiation, (\ref{eq:NEP_rad}) yields a relation that depends on the detector area $\sqrt{A}$ \cite{schmid2023fundamentals}:
\begin{equation}
    NEP=\sqrt{\frac{16\sigma_B k_B T_0^5 A}{\epsilon}},
    \label{eq:NEP_sqrt}
\end{equation}
which becomes size independent when written as specific detectivity \cite{schmid2023fundamentals}:
\begin{equation}
    D^*=\sqrt{\frac{\epsilon}{16\sigma_B k_B T_0^5}}.
    \label{eq:D*_indipendent}
\end{equation}
For an ideal infrared detector (a blackbody with absorptance $\alpha_{bb} = \epsilon_{bb} = 1$) thermally coupled to its environment on only one side, this reduces to the well-known theoretical sensitivity limit for room-temperature detectors,($T_0=290$~K and $\epsilon=1$) $D^*_{lim} = \SI{2.0e10}{\centi\meter\sqrt\hertz\per\watt}$. In the present work, we consider a resonator coupled on both sides, which slightly lowers this limit to $D^*_{bb} \approx \SI{1.4e10}{\centi\meter\sqrt\hertz\per\watt}$. Furthermore, we consider a broadband absorber with a 50\% absorptance, for which the limit reduces to ($T_0=290$~K and $\epsilon=0.5$) $D^*_{50\%} = \SI{1.0e10}{\centi\meter\sqrt\hertz\per\watt}$.

A high, uniform, and broadband absorption is advantageous for many applications and, in particular, for detectors for IR spectroscopy. Conversely, a narrowband absorption can be attractive for applications that require spectral selectivity for gas sensing or for IR imaging in the long-wavelength IR range $\SIrange{9}{10}{\micro\m}$ \cite{zhao2002optomechanical, senesac2003ir, gokhale2015low}. SiN offers a high absorption, with a peak absorption between $\lambda_{SiN}=\SI{10}{\micro\m}$ and $\lambda_{SiN}=\SI{12}{\micro\m}$ in combination with an otherwise low emissivity $\epsilon_{SiN}$. This combination renders it more responsive in narrow-band applications when compared to coated detectors. For this reason, first, we apply the analytical model to study pure SiN nanomechanical resonators without a dedicated absorber.

For sections \ref{sec:Lt}, \ref{sec:w} and \ref{sec:stress}, where we assume a \SI{50}{\nano\m} thick layer, we calculate the absorptance to be $\alpha_{SiN}(\lambda_{SiN}) = 0.255 \pm 0.007$ at the wavelength $\lambda_{SiN} = \SI{12.02}{\micro\m}$, where this thickness exhibits an absorption peak. Similarly, we determine an emissivity value of $\epsilon_{SiN} = 0.056 \pm 0.002$. Details on the derivation and the procedure used to obtain the SiN absorption spectrum at this thickness are provided in the SM. In Section \ref{sec:h}, where the thickness is varied, different values of $\alpha_{SiN}(\lambda_{SiN})$, $\lambda_{SiN}$ and $\epsilon_{SiN}$ are assumed accordingly (see Figure \ref{fig:abs_emi}). In each scenario presented in the sections \ref{sec:Lt}, \ref{sec:w}, \ref{sec:h} and \ref{sec:stress}, we also explore the case of an IR absorber layer with an absorptance $\alpha_{abs}=50\%$ deposited onto the SiN detector, allowing for broader absorption. Throughout the analysis, a fundamental trade-off becomes evident: a higher specific detectivity confined to a narrow spectral band, versus a lower detectivity extended over a broad wavelength range, each corresponding to different application domains.

\section{Geometry ($\mathrm{L}$, $\mathrm{L_t}$, $\mathrm{h}$, and $\mathrm{w}$)}

\label{sec:geometry}
\begin{table}
    \centering
    \begin{tabular}{c|c|c|c|c|c}
         $\rho\ [kg/m^3]$ & $E\ [Pa]$ & $\nu$ & $\alpha_{th}\ [K^{-1}]$ & $\kappa\ [W/(mK)]$ & $c_p\ [J/(kgK)]$ \\\hline\hline
         $3000$ & $250\times10^9$ & $0.27$ & $1.5\times10^{-6}$ & $3$ & $700$ \\
    \end{tabular}
    \caption{Silicon nitride parameters used. From left to right: mass density, Young's modulus, Poisson ratio, thermal expansion coefficient, thermal conductivity, and specific heat capacity.}
    \label{tab:mat_par}
\end{table}

Geometry is typically the first aspect to consider when designing a nanomechanical IR detector, and it is the focus of this section. The analysis begins with the parameters specific to trampoline designs, namely $L_t$ and $w$, followed by the thickness $h$, which is shared by both drumheads and trampolines. For all cases, we present results across five different values of the detection area side length $L$. The impact of each geometrical variable is evaluated with respect to the thermal time constant ($\tau_{th}$), noise equivalent power (NEP), and specific detectivity ($D^*$). 
Table \ref{tab:mat_par} lists the SiN parameters considered for this work \cite{ piller2022thermal, kanellopulos2024stress, toivola2003influence, ftouni2015thermal}.

\subsection{Tethers length, $L_t$}
\label{sec:Lt}

\begin{figure}
    \centering
    \includegraphics[width=1\linewidth]{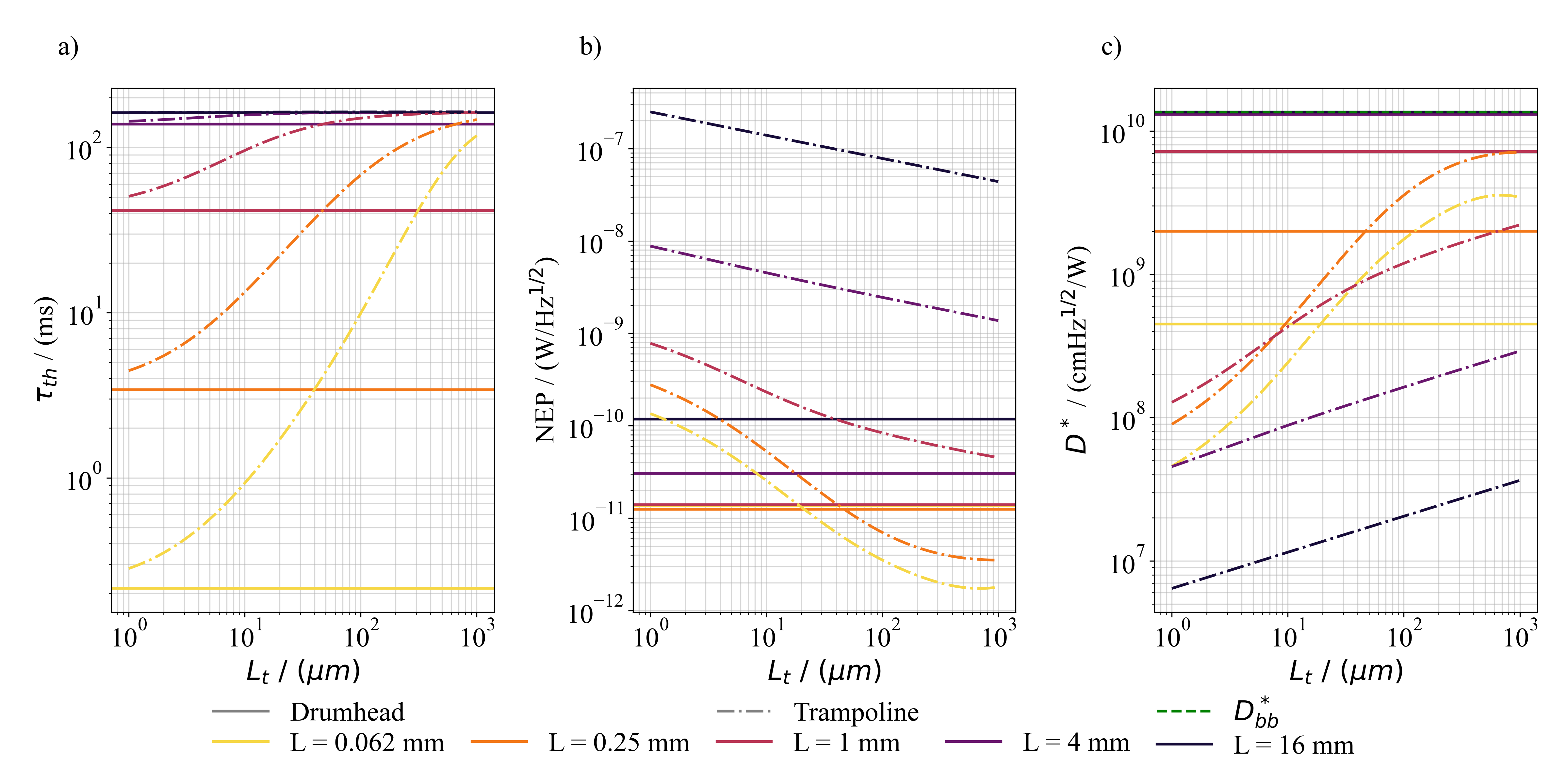}
    \caption{Influence of the tether length $L_t$ on the figures of merit: a) thermal time constant $\tau_{th}$, b) noise equivalent power NEP, and c) specific detectivity $D^*$ at $\lambda_{SiN} = \SI{11.86}{\micro\m}$ for plain SiN resonators without a dedicated IR absorber. The dashed black line in c) marks the room-temperature sensitivity limit. For all panels, the stress is fixed at $\sigma = \SI{200}{\mega\pascal}$, the thickness $h = \SI{50}{\nano\m}$, and the tether width at $w = \SI{5}{\micro\m}$.}
    \label{fig:FOM_Lt}
\end{figure}

First, we examine how the tether length, $L_t$, influences the performance of trampolines with different detection areas. The results of this analysis for bare SiN are shown in Figure \ref{fig:FOM_Lt}. In this section, the tethers width is fixed at $w=\SI{5}{\micro\m}$. For comparison, in all panels of Figure \ref{fig:FOM_Lt}, we display the values corresponding to drumheads of the same thickness and stress, as horizontal solid lines because they are independent of $L_t$. For all the designs, the thickness and tensile stress are set to $h=\SI{50}{\nano\m}$ and $\sigma=\SI{200}{\mega\pascal}$, respectively. 

The effect of $L_t$ on the thermal conductance $G$ is twofold, as shown in Eq. (\ref{eq:G_trampoline}): on the one hand, longer tethers reduce heat transfer via conduction; and on the other hand, increasing $L_t$ enlarges the total radiating surface, enhancing heat loss through radiation. Additionally, the total heat capacitance $C$ increases linearly with $L_t$, as indicated in Eq. (\ref{eq:C_trmapoline}). The combined effect of these contributions is illustrated in Figure \ref{fig:FOM_Lt}a, which shows $\tau_{th}$ as a function of $L_t$. For small $L_t$, heat from the center of the detection area is primarily dissipated through conduction along the tethers (high $G_{cond}$). As $L_t$ increases, the contribution of $G_{cond}$ diminishes, and the radiative contribution ($G_{rad}$) becomes significant, coupling the resonator more strongly to the environment via radiation. As a result, $\tau_{th}$ increases until asymptotically reaching the radiative heat transfer plateau.  

The NEP, shown in Figure~\ref{fig:FOM_Lt}b, exhibits a clear dependence on the tether length $L_t$. The most sensitive detector corresponds to the smallest trampoline geometry, with a minimum NEP occurring around $L_t\simeq\SI{700}{\micro\m}$. 

Finally, the specific detectivity as a function of $L_t$ is shown in Figure \ref{fig:FOM_Lt}c. In this case, the trampoline with detection area side length $L=\SI{0.25}{\milli\m}$ exhibits the highest $D^*$. However, even at the peak of its curve, corresponding to a tether length of $L_t\simeq\SI{1}{\milli\m}$, its $D^*$ is comparable to that of the drumhead with side length $L=\SI{1}{\milli\m}$. For this figure of merit, the overall best-performing devices are drumheads with $L=\SIlist{4;16}{\milli\m}$, both reaching the room-temperature detectivity limit $D^*_{bb}$. This result underscores the crucial role of the detection area ($A$ in equation (\ref{eq:D*})): although smaller resonators exhibit lower NEP, this advantage is outweighed by the larger IR collection area of bigger structures.

\begin{figure}
    \centering
    \includegraphics[width=1\textwidth]{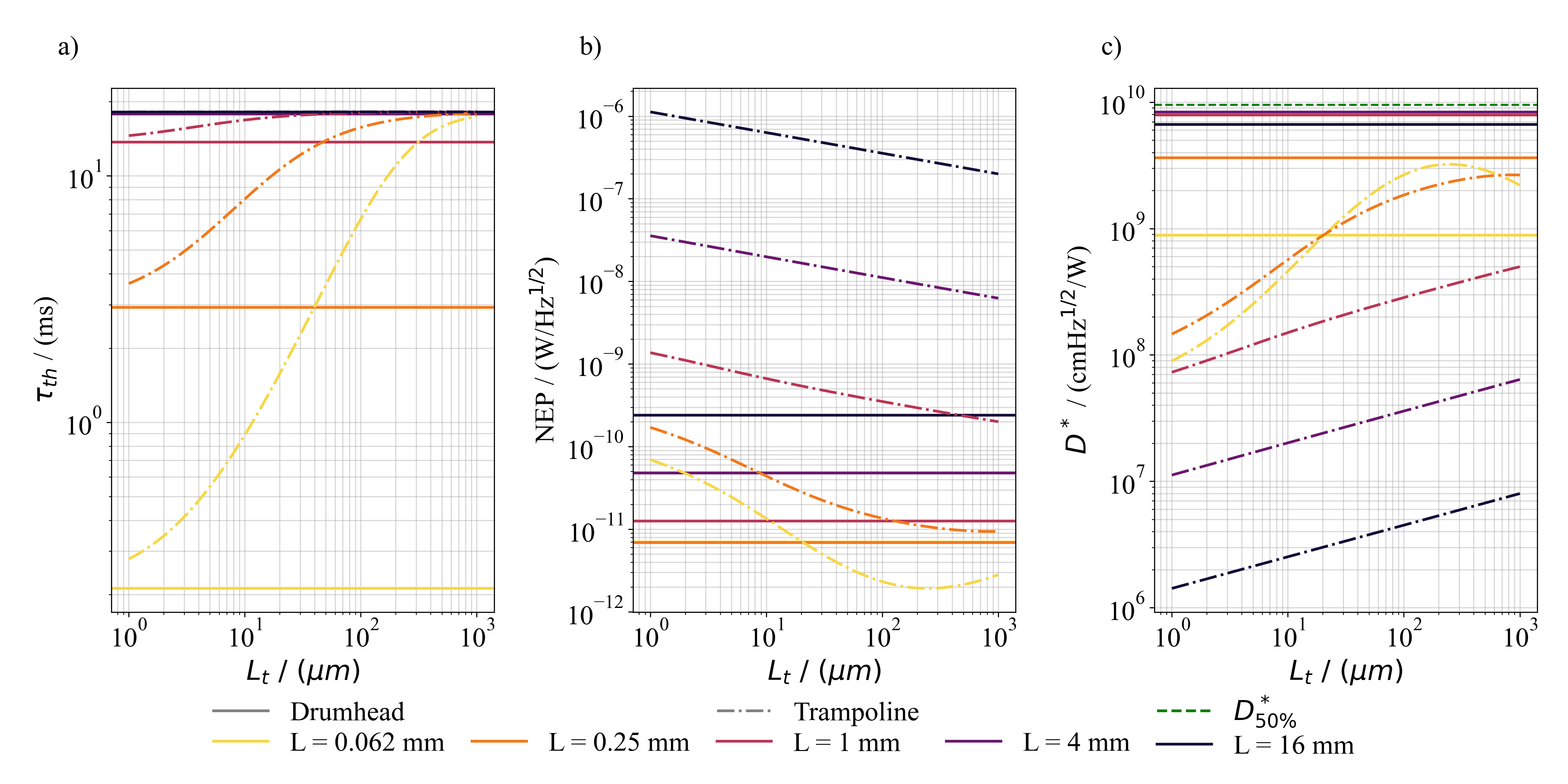}
    \caption{Influence of the tether length $L_t$ on the figures of merit: a) thermal time constant $\tau_{th}$, b) noise equivalent power NEP, and c) specific detectivity $D^*$ for a SiN resonator with an IR absorber. The dashed black line in c) marks the room-temperature sensitivity limit for $\alpha_{abs}=50\%$. For all panels, the stress is fixed at $\sigma = \SI{200}{\mega\pascal}$, the thickness $h = \SI{50}{\nano\m}$, and the tether width at $w = \SI{5}{\micro\m}$.}
    \label{fig:3in1_Lt_50abs}
\end{figure}
So far, we have evaluated the performance of bare SiN resonators at the fixed wavelength $\lambda_{SiN} = \SI{12.02}{\micro\m}$, corresponding to the measured peak in absorptance. All previous results are therefore specific to this wavelength. However, in the field of infrared detection, it is often desirable to develop broadband detectors. One promising approach is the deposition of a metal layer on a dielectric, which can achieve absorptance values up to $\alpha_{abs} = 50\%$ across a broad spectral range \cite{hilsum1954absorption}. Recent studies have demonstrated such high and nearly flat absorption spectra using platinum or gold thin films deposited on SiN \cite{piller2022thermal, martini2025uncooled, luhmann2020ultrathin}. As shown in those works, the addition of a thin metal layer does not significantly affect the mechanical properties of the resonator, although it does influences its thermal and optical behaviour, and provides an almost flat and wavelength independent absorption spectrum, over the infrared range, up to the THz region. Now we investigate the impact of adding an absorber with $\alpha_{abs} = \epsilon_{abs} = 50\%$ on the SiN resonator, under the assumption that the mechanical properties of SiN remain largely unchanged. 

The results of this analysis for varying tethers length $L_t$ are shown in Figure \ref{fig:3in1_Lt_50abs}. As observed for bare SiN (Figure \ref{fig:FOM_Lt}b), the best NEP values are obtained for the smallest trampoline (Figure \ref{fig:3in1_Lt_50abs}b). However, in this case, the lowest NEP occurs around $L_t \simeq \SI{250}{\micro\m}$. In general, smaller structures outperform larger ones, both for trampolines and drumheads. 
Overall, we observe that adding an absorbing layer lowers $\tau_{th}$ (Figure \ref{fig:3in1_Lt_50abs}a) and shifts the optimal designs toward smaller resonators (Figure \ref{fig:3in1_Lt_50abs}c) compared to the bare SiN scenario. Although the absolute values of the figures of merit are slightly lower than in the bare SiN case, it is important to emphasize that, with the absorber, these performance levels are maintained across the entire MIR spectral range, whereas previously they were limited to the specific peak wavelength of SiN.

\subsection{Tethers width, $w$}
\label{sec:w}

The resulting behaviour of NEP and $D^*$ as a function of $w$ closely mirrors that observed for $L_t$. To avoid redundancy, we omit a detailed discussion here; the full analysis is provided in the SI, both for the case of bare SiN and with the metal absorber.

\subsection{Thickness, $h$}
\label{sec:h}

\begin{figure}
    \centering
    \includegraphics[width=0.51\linewidth]{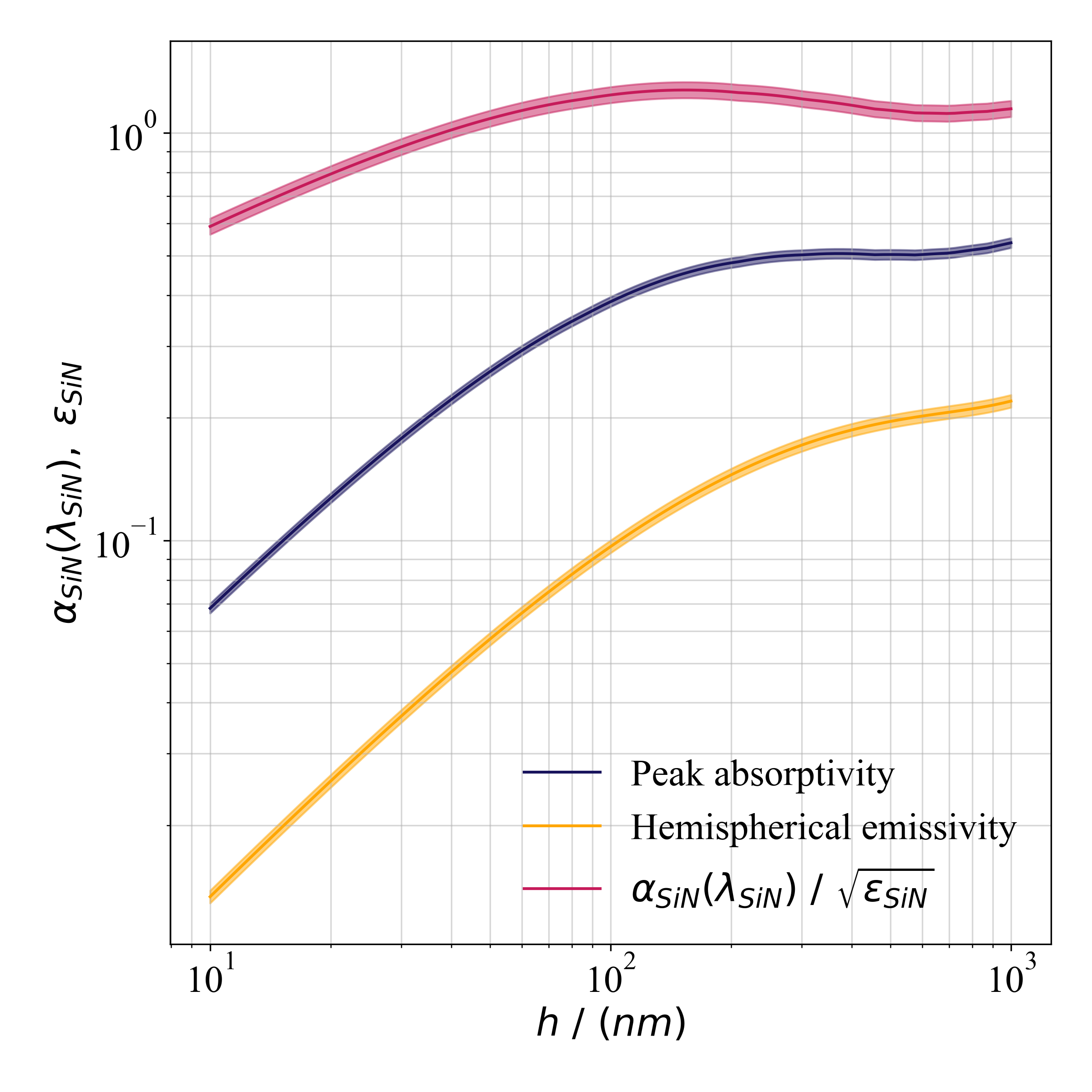}
    \caption{Effect of the thickness $h$ over SiN absorptance at its absorption peak $\lambda_{SiN}=\SI{11.86}{\micro\m}$ (blue line) and the hemispherical emissivity (orange line). Both grow as the SiN gets thicker. $\alpha_{SiN}(\lambda_{SiN})$ has a maximum for $h\eqsim\SI{250}{\nano\m}$ and later shows a local minimum for $h\eqsim\SI{700}{\nano\m}$. The green line is the ratio $\alpha(\lambda_{SiN})/\sqrt{\epsilon_{SiN}}$, and it has a maximum for $h=\SI{151}{\nano\m}$.}
    \label{fig:abs_emi}
\end{figure}

In nanomechanical resonators, the thickness can range from a few to several hundred \SI{}{\nano\m}, and in some cases, up to a few \SI{}{\micro\m}. In this work, we consider the range $h = \SI{10}{\nano\m}$ to $\SI{1}{\micro\m}$. In the thin-film regime, variations in thickness significantly affect the film’s optical properties, such as absorptance $\alpha(\lambda)$, intended as the fraction of the absorbed light at one specific wavelength, and emissivity, or hemispherical emissivity $\epsilon$, which is the efficiency with which the film radiates energy compared to that of a blackbody at the same temperature, integrated over all emission angles and wavelengths \cite{edalatpour2013size, bergman2011fundamentals}. For each value of $h$, we computed $\alpha_{SiN}(\lambda_{SiN})$ and $\epsilon_{SiN}$, taking into account possible interference effects with the thin film \cite{kanellopulos2024stress}. The result is shown in Figure \ref{fig:abs_emi}, where the blue line represents the SiN peak absorptance ($\alpha_{SiN}(\lambda_{SiN})$), the orange line the hemispherical emissivity ($\epsilon_{SiN}$), and the green line the ratio $\alpha(\lambda_{SiN})/\sqrt{\epsilon_{SiN}}$. This latter quantity is a strong indicator of the SiN performance: on one hand, a thicker material increases the absorbed power, thereby enhancing the power responsivity (equation (\ref{eq:Rp})); on the other hand, a lower emissivity reduces the radiative coupling to the environment, which is also desirable. Therefore, rather than solely maximizing the absorptance or minimizing the emissivity, the optimal strategy is to maximize the ratio. The term $\alpha(\lambda)/\sqrt{\epsilon}$ can be seen as a specific detectivity enhancement factor typical of metamaterials, for which, in a very narrow spectral region, it holds true that $\alpha(\lambda)>\epsilon$. Figure \ref{fig:abs_emi} shows that for SiN the maximum of the ratio occurs at $h=\SI{151}{\nano\m}$. For more details on how the absorptance and emissivity curves are obtained, the reader is referred to the SM. 
\begin{figure}
    \centering
    \includegraphics[width=1\linewidth]{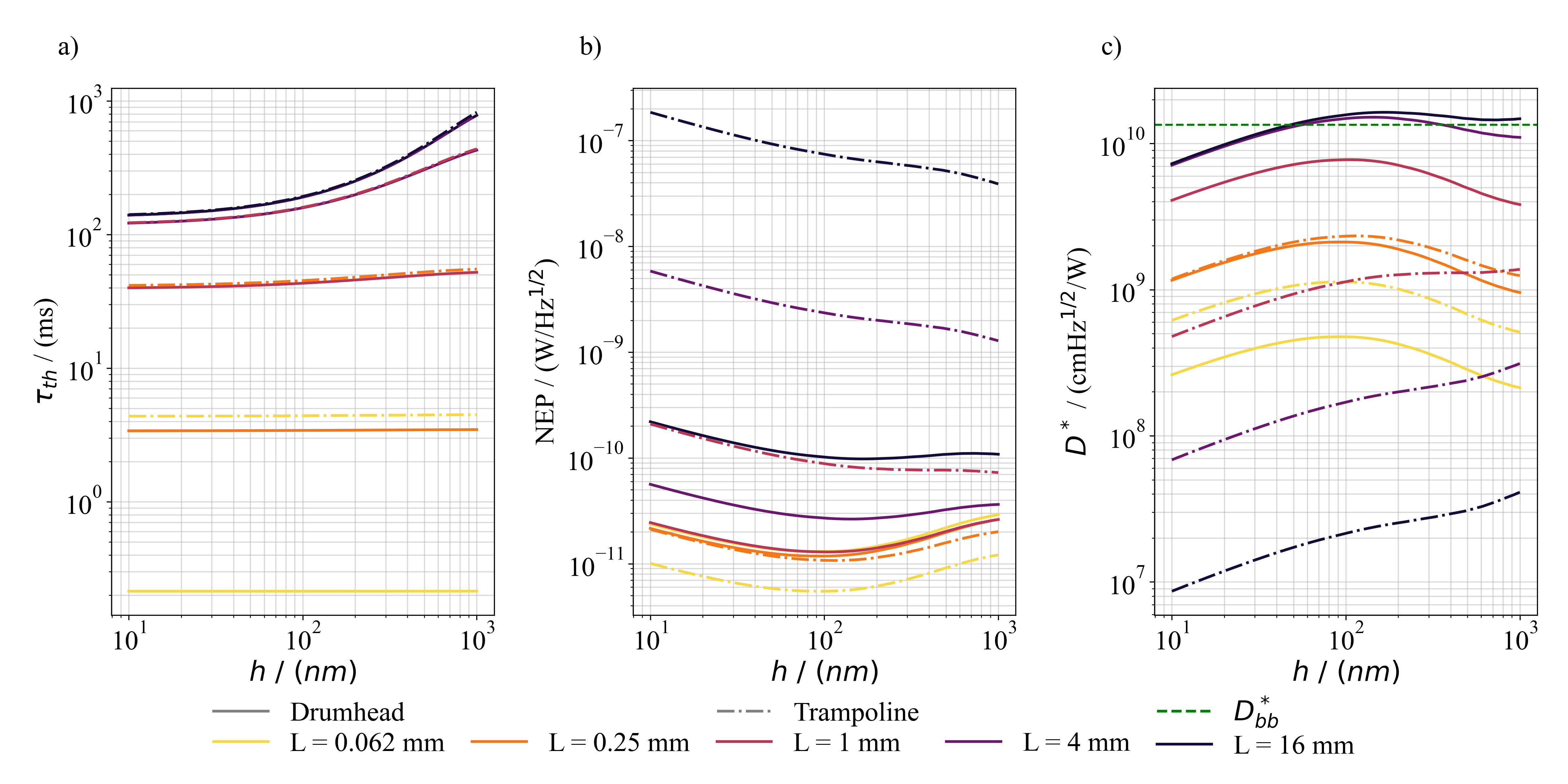}
    \caption{Influence of the thickness $h$ on the figures of merit: a) thermal time constant $\tau_{th}$, b) noise equivalent power (NEP), and c) specific detectivity $D^*$ at $\lambda_{SiN} = \SI{11.86}{\micro\m}$ for plain SiN resonators without a dedicated IR absorber. The dashed black line in c) marks the room-temperature sensitivity limit. For all panels, the stress is fixed at $\sigma = \SI{200}{\mega\pascal}$, the tether length at $L_t = \SI{50}{\micro\m}$, and the tether width at $w = \SI{5}{\micro\m}$.}
    \label{fig:FOM_h}
\end{figure}

During the analysis of thickness dependence, all other parameters are kept constant: the stress is fixed at $\sigma = \SI{200}{MPa}$, and for the trampoline geometries, the tether length and width are set to $L_t = \SI{50}{\micro\m}$ and $w = \SI{5}{\micro\m}$, respectively. According to equations (\ref{eq:C_drumhead}) and (\ref{eq:C_trmapoline}), increasing $h$ enhances both the heat capacity $C$ and the thermal conductance $G$. For small drumheads ($L = \SI{0.062}{\milli\m}$ and $\SI{0.25}{\milli\m}$), where conduction dominates, the $h$-dependence in $C$ and $G$ effectively cancels out, yielding a nearly constant $\tau_{th}$ (see figure \ref{fig:FOM_h}a). In contrast, for larger drumheads ($L = \SIlist{4;16}{\milli\m}$) the radiative term becomes significant, causing $\tau_{th}$ to increase with $h$. A similar trend is observed for trampolines, where the tethers, by hindering conduction and augmenting the radiating surface, make the influence of $h$ on $\tau_{th}$ evident even for $L = \SI{0.25}{\milli\m}$. Thus, the presence of tethers enhances thermal isolation, rendering trampolines slower than drumheads for a given $L$.

Figure \ref{fig:FOM_h}b shows the results for the NEP. The most sensitive device is the trampoline with $L=\SI{0.062}{\milli\m}$, with the lowest values obtained for $h\simeq\SI{100}{\nano\m}$. The slight deviation in thickness from the value corresponding to the maximum of the green curve in Figure \ref{fig:abs_emi} arises from the non-negligible contribution of thermal conduction, which becomes increasingly significant as the resonator thickness grows. 

Finally, the model predictions for the specific detectivity $D^*$ are presented in Figure \ref{fig:FOM_h}c. In this case, the best performance is achieved by drumheads, with the two designs featuring the largest detection areas yielding the highest overall values. For these configurations, the predicted $D^*$ values at thicknesses around $\SI{200}{\nano\m}$ surpasses the fundamental room-temperature limit, $D^*_{bb}$, indicated by the green dotted line. This somehow counterintuitive result is in line with what is shown in Figure \ref{fig:abs_emi}, and demonstrates that bare SiN, when probed at its absorptance peak wavelength $\lambda_{SiN}$, acts similarly to a metamaterial, showing $\alpha_{SiN}(\lambda_{SiN}) > \epsilon_{SiN}$. Overall, drumheads outperform trampolines of the same size, with the exception of the smallest drumhead, which is outperformed by the trampoline with $L = \SI{0.062}{\milli\m}$.

\begin{figure}
    \centering
    \includegraphics[width=1\linewidth]{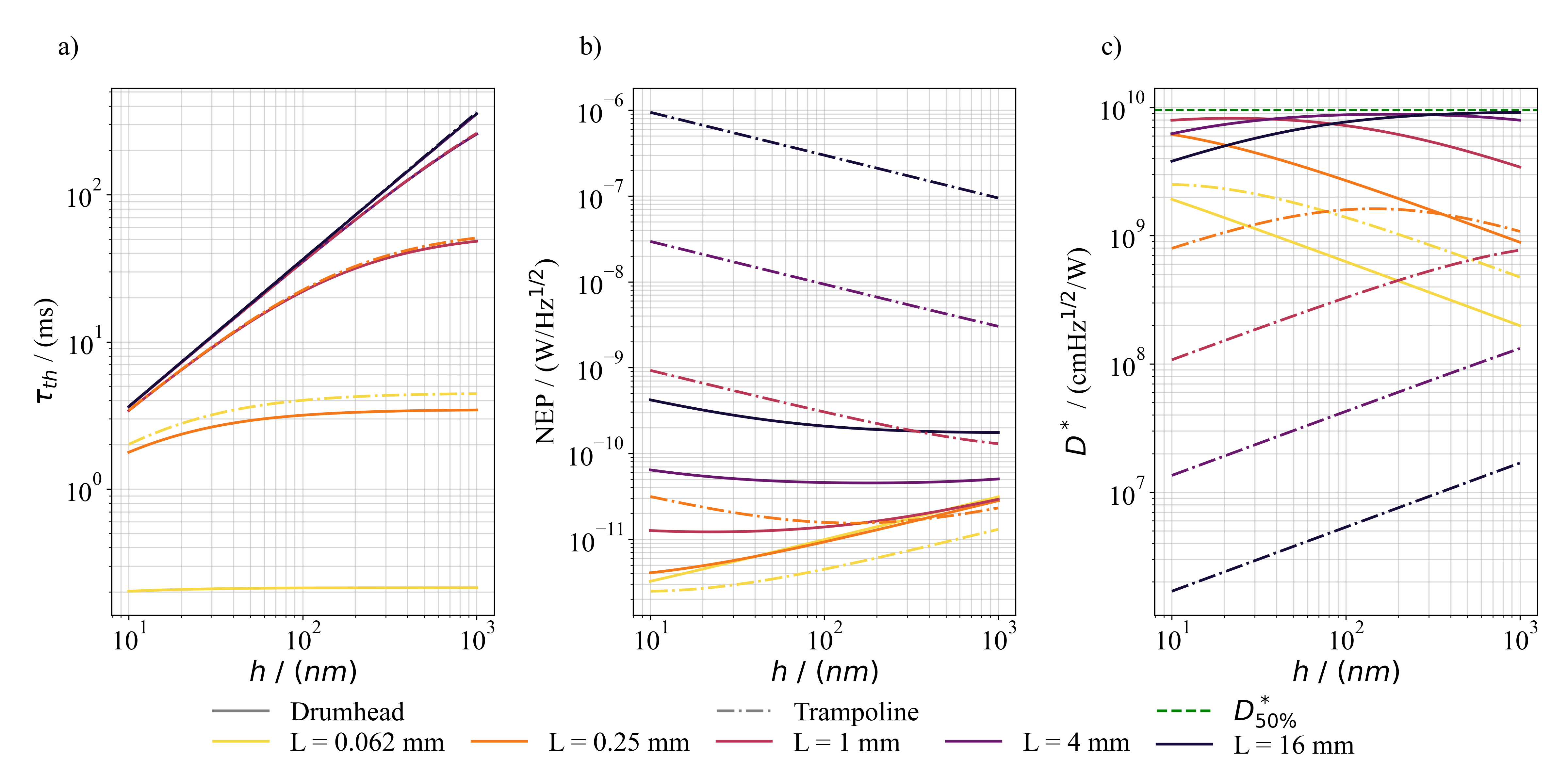}
    \caption{Influence of the thickness $h$ on the figures of merit: a) thermal time constant $\tau_{th}$, b) noise equivalent power (NEP), and c) specific detectivity $D^*$ for SiN resonators with a dedicated IR absorber. The dashed black line in c) marks the room-temperature sensitivity limit for $\alpha_{abs}=50\%$. For all panels, the stress is fixed at $\sigma = \SI{200}{\mega\pascal}$, the tether length at $L_t = \SI{50}{\micro\m}$, and the tether width at $w = \SI{5}{\micro\m}$.}
    \label{fig:FOM_h_abs50}
\end{figure}
The presence of the metal layer which gives the detector a constant absorptance and emissivity of $50\%$ can be noticed in the results shown in Figure \ref{fig:FOM_h_abs50}. Because of this, the influence of $h$ is now limited to heat transfer through conduction, $G_{cond}$, for both trampolines and drumheads, with no impact on $G_{rad}$. In Figure \ref{fig:FOM_h_abs50}a, we notice the thermal time constant significantly decreasing for every design and  when compared to the bare SiN scenario (Figure \ref{fig:FOM_h}a). In Figure \ref{fig:FOM_h_abs50}b, we see the results for the NEP, with smaller detectors exhibiting superior performance, particularly at reduced thicknesses. 
As the lateral dimensions of the structures increase, the optimal NEP shifts toward larger thickness values. This trend arises because a larger $L$ enhances radiative exchange, so the sweetspot between radiative losses and thermal conduction is reached at greater thicknesses. A similar trend is observed for $D^*$ (Figure \ref{fig:FOM_h_abs50}c), for both drumheads and trampolines. This behaviour reflects the fact that increasing detector area requires thicker substrates to achieve the optimal balance between $G_{cond}$ and $G_{rad}$.

For all the geometrical variables, the results for the power responsivity $R_P$ are presented in the SM.

\section{Tensile stress, $\sigma$}
\label{sec:stress}

\begin{figure}
    \centering
    \includegraphics[width=1\linewidth]{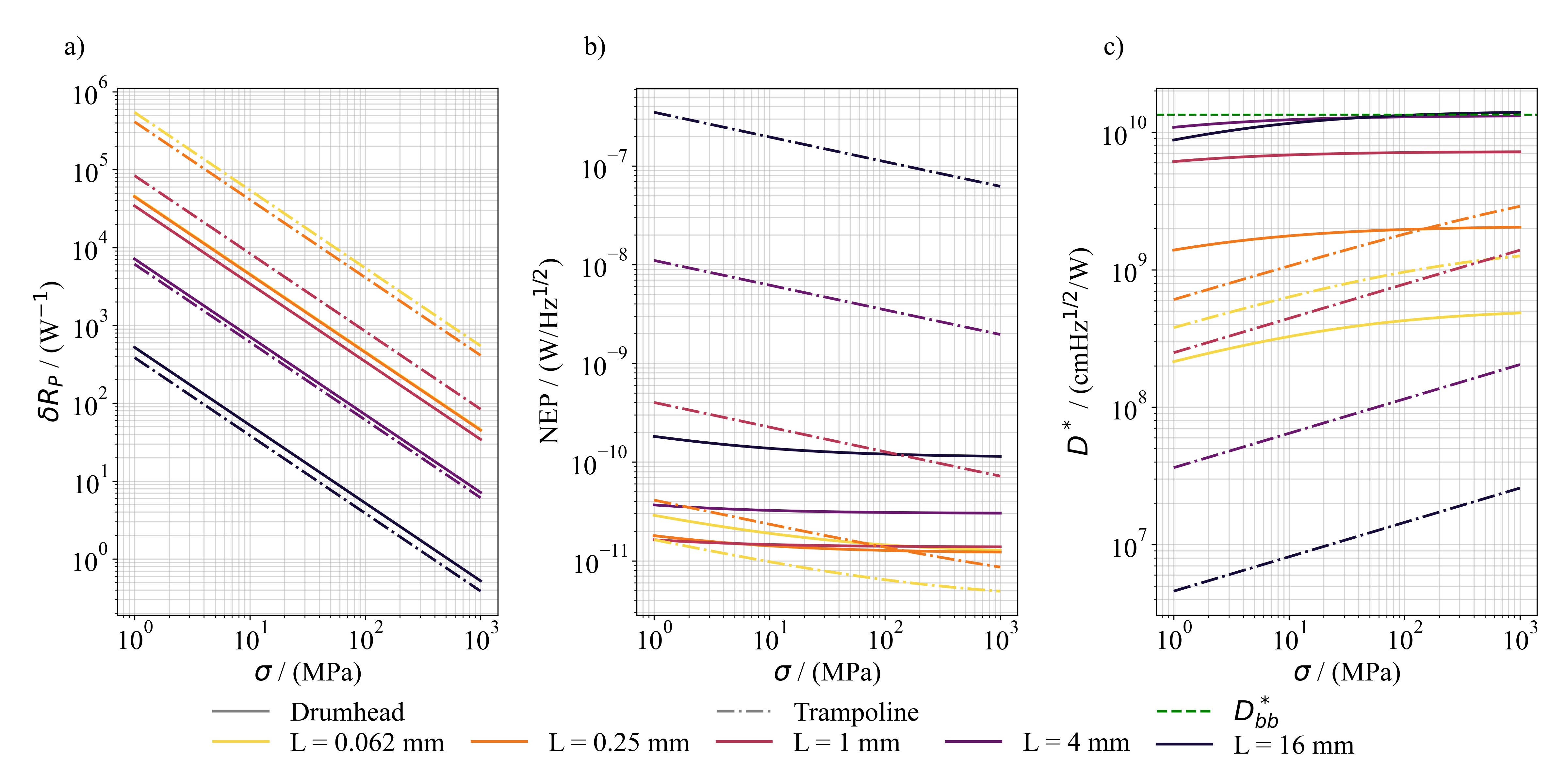}
    \caption{Influence of the tensile stress $\sigma$ on the figures of merit: a) power responsivity $R_P$; b) noise equivalent power (NEP); c) specific detectivity $D^*$ at $\lambda_{SiN} = \SI{11.86}{\micro\m}$ for plain SiN resonators without a dedicated IR absorber. The dashed black line in c) marks the room-temperature sensitivity limit. For all panels, the thickness is fixed at $h = \SI{50}{\nano\m}$, the tether length at $L_t = \SI{50}{\micro\m}$, and the tether width at $w = \SI{5}{\micro\m}$.}
    \label{fig:FOM_sigma}
\end{figure}
This section is dedicated to the analysis of the influence of tensile stress $\sigma$ on the detectors' performance. Often referred to as prestress, it is to some extent a tunable parameter of mechanical resonators. Different deposition methods and parameters result in varying stress conditions in the deposited material. From Section \ref{sec:FOM_tau}, we know that $\sigma$ has no influence on either the thermal conductance $G$ or the heat capacity $C$ (see equations (\ref{eq:C_drumhead}) and (\ref{eq:G_drumhead}) for drumheads, and equations (\ref{eq:C_trmapoline}) and (\ref{eq:G_trampoline}) for trampolines); thus, it does not affect the thermal time constant of the device. In some cases, differences in deposition techniques result in variations in crystallinity (e.g., $\alpha$-phase SiN vs. $\beta$-phase SiN vs. amorphous SiN). These differences typically entail changes in material properties, such as specific heat capacity and thermal conductivity, which would influence $\tau_{th}$. While the authors are aware of this, the investigation of such effects falls outside the scope of the present study. Here, we consider the case of amorphous SiN deposited with LPCVD, for which we assume that prestress conditions do not affect the thermal properties of the material. Consequently, in presenting the model results for the various figures of merit, we exclude $\tau_{th}$. Analysing the effects of stress by comparing drumheads to trampolines is non-trivial. This is due to the well-known influence of the trampoline geometry on the stress distribution: the stress is significantly increased in the tethers and reduced in the central area compared to the nominal stress $\sigma_0$ of the deposited material, as shown through finite element (FEM) simulations in \cite{piller2022thermal}. Nevertheless, the results presented in this section remain generally valid when considering the value of the stress $\sigma$ in the specific region of the detector where the IR light impinges and is absorbed. For the following study on varying the stress in the range $\sigma=\SIrange{1}{1000}{\mega\pascal}$, the thickness of all structures is fixed to $h=\SI{50}{\nano\m}$, and for the trampolines, the tethers have width $w=\SI{5}{\micro\m}$, and length $L_t=\SI{50}{\micro\m}$.

The dependency of $R_P$ on stress arises from the linear dependence of the temperature responsivity $R_T$ (equations (\ref{eq:Rt_drumhead}) and (\ref{eq:Rt_trampoline})) on $\sigma$. For both designs, $R_T$ decreases linearly with increasing stress, and so does the power responsivity, as shown in Figure \ref{fig:FOM_sigma}a. The differences in $R_P$ values between structures of different sizes stem from the expected variations in $G$: detectors with smaller detection areas generally perform better. The results show that drumheads with $L=\SIlist{4;16}{\milli\m}$ exhibit slightly higher responsivity than trampolines with equivalent detection areas. However, for smaller $L$ values, trampolines display higher $R_P$ compared to drumheads. This improvement is attributed to their enhanced thermal isolation of the centre of the detection area, enabled by the presence of tethers that reduce conductive heat loss.

The situation is almost the opposite when it comes to the noise equivalent power. As shown in Figure \ref{fig:FOM_sigma}b, the NEP generally improves with increasing values of $\sigma$, and drumheads tend to outperform trampolines, with the sole exception of the smallest trampoline, which exhibits the lowest NEP among all designs. Both the inverse trend of NEP and the relative performance of drumheads compared to trampolines can be explained by frequency stability. As detailed in the SI, the significantly lower frequency noise of drumheads relative to trampolines, along with the general improvement in frequency stability with increased stress, compensates for the differences in power responsivity observed in Figure \ref{fig:FOM_sigma}a. Interestingly, while smaller detection areas still correspond to better performance in trampolines, for drumheads, the detectors with bigger areas are those that perform best across the entire $\sigma$ range under investigation. 

The specific detectivity results are presented in Figure \ref{fig:FOM_sigma}c. Again, we observe what is seen for all the other variables in section \ref{sec:geometry}: introducing the detection area $A$ in the calculation highlights the importance of collecting as much light as possible from the source. It is now the large drumheads which show the highest $D^*$, with $L=\SI{4}{\milli\m}$ being the top performer for $\SI{1}{\mega\pascal} \leq \sigma \leq \SI{10}{\mega\pascal}$ and matching the performance of $L=\SI{16}{\milli\m}$ for $\sigma > \SI{10}{\mega\pascal}$. For a given detection area side length $L$, drumheads always perform better than trampolines, with the case of $L=\SI{0.062}{\micro\m}$ being the only exception to this. However, it consistently holds that $D^*$ increases with higher values of tensile stress $\sigma$, independently of the detector design.

\begin{figure}
    \centering
    \includegraphics[width=1\linewidth]{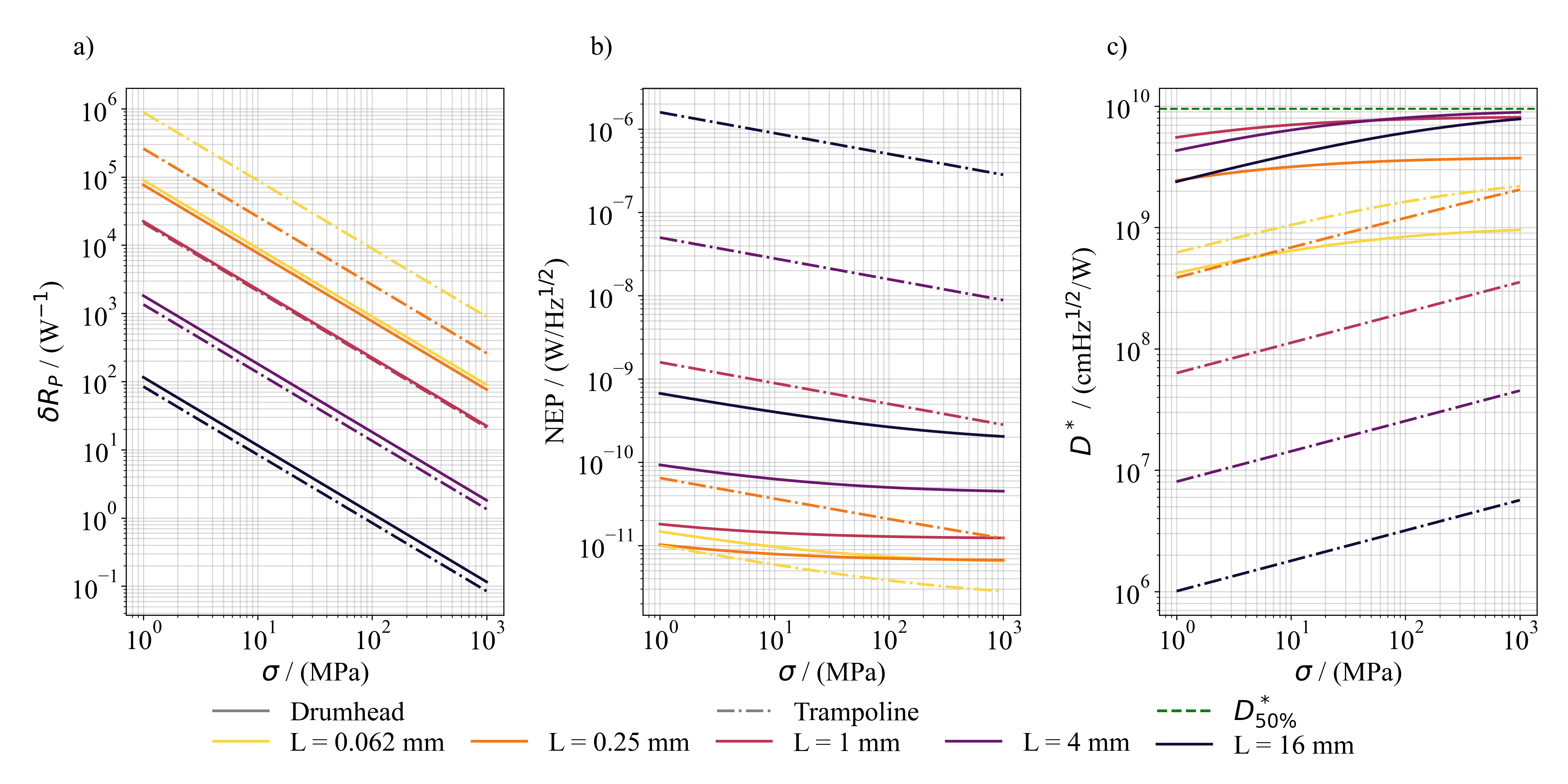}
    \caption{Influence of the tensile stress $\sigma$ on the figures of merit: a) power responsivity $R_P$; b) noise equivalent power (NEP); c) specific detectivity $D^*$ for SiN resonators with a dedicated IR absorber. The dashed black line in c) marks the room-temperature sensitivity limit for $\alpha_{abs}=50\%$. For all panels, the thickness is fixed at $h = \SI{50}{\nano\m}$, the tether length at $L_t = \SI{50}{\micro\m}$, and the tether width at $w = \SI{5}{\micro\m}$.}
    \label{fig:FOM_sigma_abs50}
\end{figure}
Very similar results are obtained with the inclusion of the metal absorber layer, as shown in Figure \ref{fig:FOM_sigma_abs50}. The overall trends observed for bare SiN are preserved in this case as well. A slight increase in power responsivity is observed (Figure \ref{fig:FOM_sigma_abs50}a), accompanied by a modest reduction in performance with respect to NEP and $D^*$ (Figures \ref{fig:FOM_sigma_abs50}b and \ref{fig:FOM_sigma_abs50}c, respectively). 

\begin{figure}
    \centering
    \includegraphics[width=1\linewidth]{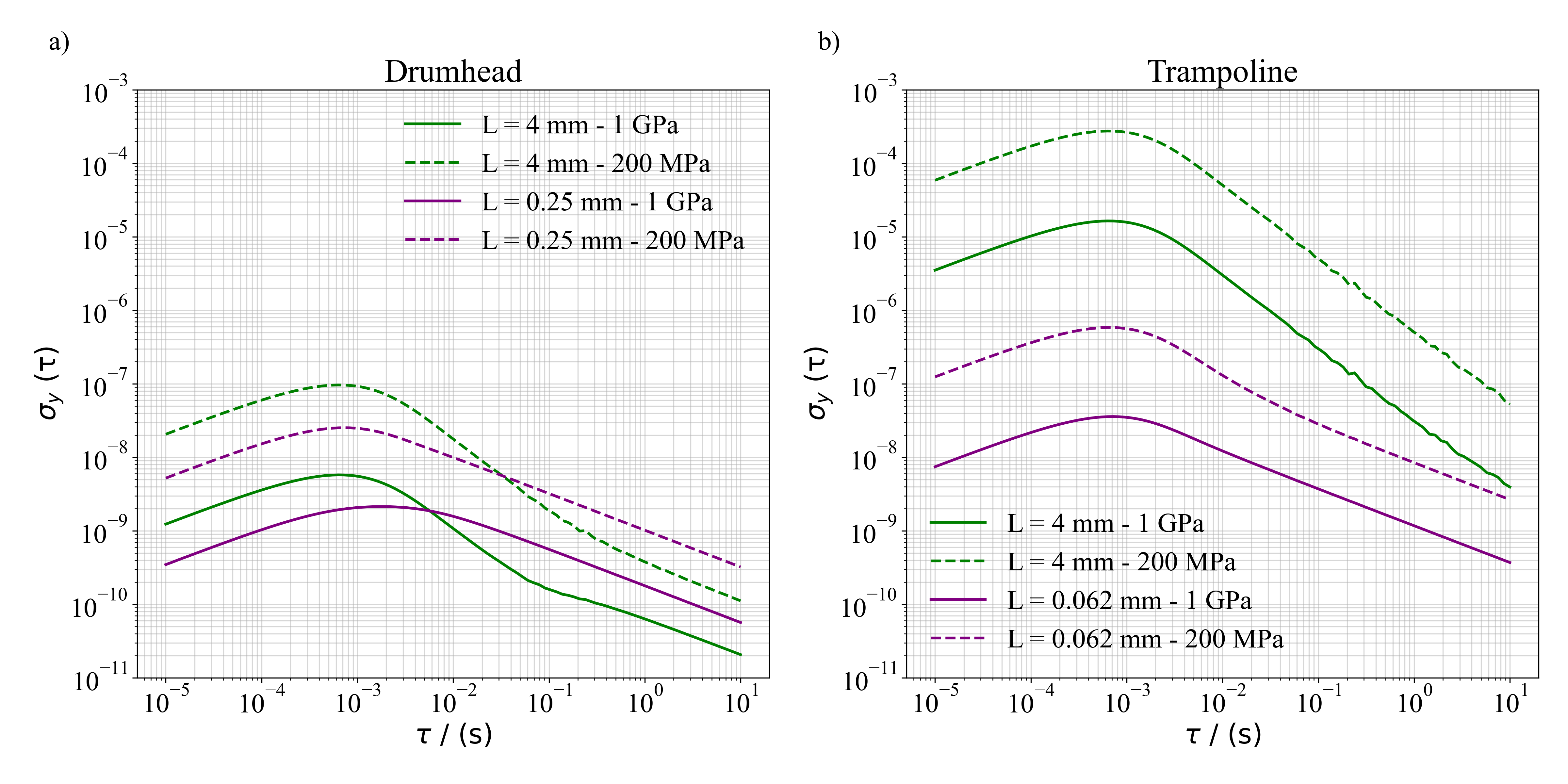}
    \caption{Theoretical ADs, obtained by adding together thermomechanical and detection noise, for a) drumheads and b) trampolines, for different $\sigma$. For this study, the thickness is set to $h=\SI{50}{\nano\m}$ while for trampolines the tether length is $L_t = \SI{50}{\micro\m}$ and tether width is $w = \SI{5}{\micro\m}$.}
    \label{fig:ADs for sigma}
\end{figure}

The key takeaway from Figure~\ref{fig:FOM_sigma} and \ref{fig:FOM_sigma_abs50} is that higher stress leads to improved detector performance. However, the practical implications of employing high-stress resonators for IR detection must also be considered. The enhanced NEP and $D^*$ observed at higher stress levels originate from the overall improvement in frequency stability associated with such resonators. This is illustrated in Figure \ref{fig:ADs for sigma}, where we plot the theoretical Allan deviations, accounting for both additive and temperature fluctuation noise, as a function of integration time $\tau$, for drumhead and trampoline resonators of different sizes and two commonly used stress values: \SI{200}{\mega\pascal} and \SI{1}{\giga\pascal}. The analysis assumes a closed-loop measurement with a phase-locked loop (PLL) bandwidth of \SI{100}{\hertz}, and a thermomechanical-to-detection noise ratio $\mathcal{K} = 0.01$ for a typical optical interferometer that has a displacement sensitivity in the order of \SI{1}{\pico\m\per\sqrt\hertz}. In all cases, the high-stress resonators (solid lines) exhibit lower noise than their low-stress counterparts (dashed lines). Notably, for the most promising designs, namely, the smaller trampolines (purple lines in figure \ref{fig:ADs for sigma}b), which exhibit the lowest NEP in Figure \ref{fig:FOM_sigma}b, and the larger membranes (solid green line in figure \ref{fig:ADs for sigma}a), which show the highest $D^*$ in Figure \ref{fig:FOM_sigma}c, the Allan deviation are below $10^{-7}$, and below $10^{-9}$, respectively. At such low levels, the detector performance is increasingly limited by other noise sources, such as absorbtion-desorption noise \cite{cleland2002noise, djuric2002adsorption}, defect motion \cite{fong2012frequency}, random walk frequency drift, and photothermal back-action noise \cite{kanellopulos2025comparative, martini2025uncooled, zhang2023demonstration}. Ultimately, the precision of such measurements is limited by the frequency stability of the reference oscillators in the readout electronics — typically oven-controlled crystal oscillators (OCXOs) — which offer stabilities on the order of $10^{-9}$. In practice, these additional noise sources will likely dominate in high-stress, frequency-stable resonators, making the low values shown in Figure~\ref{fig:ADs for sigma} practically unachievable. It may therefore be more practical to employ a detector with moderately lower stress $\sigma$, which offers slightly reduced frequency stability but benefits from a higher $R_P$.

\section{Discussion}
\label{discussion}

\begin{figure}
    \centering
    \includegraphics[width=1\linewidth]{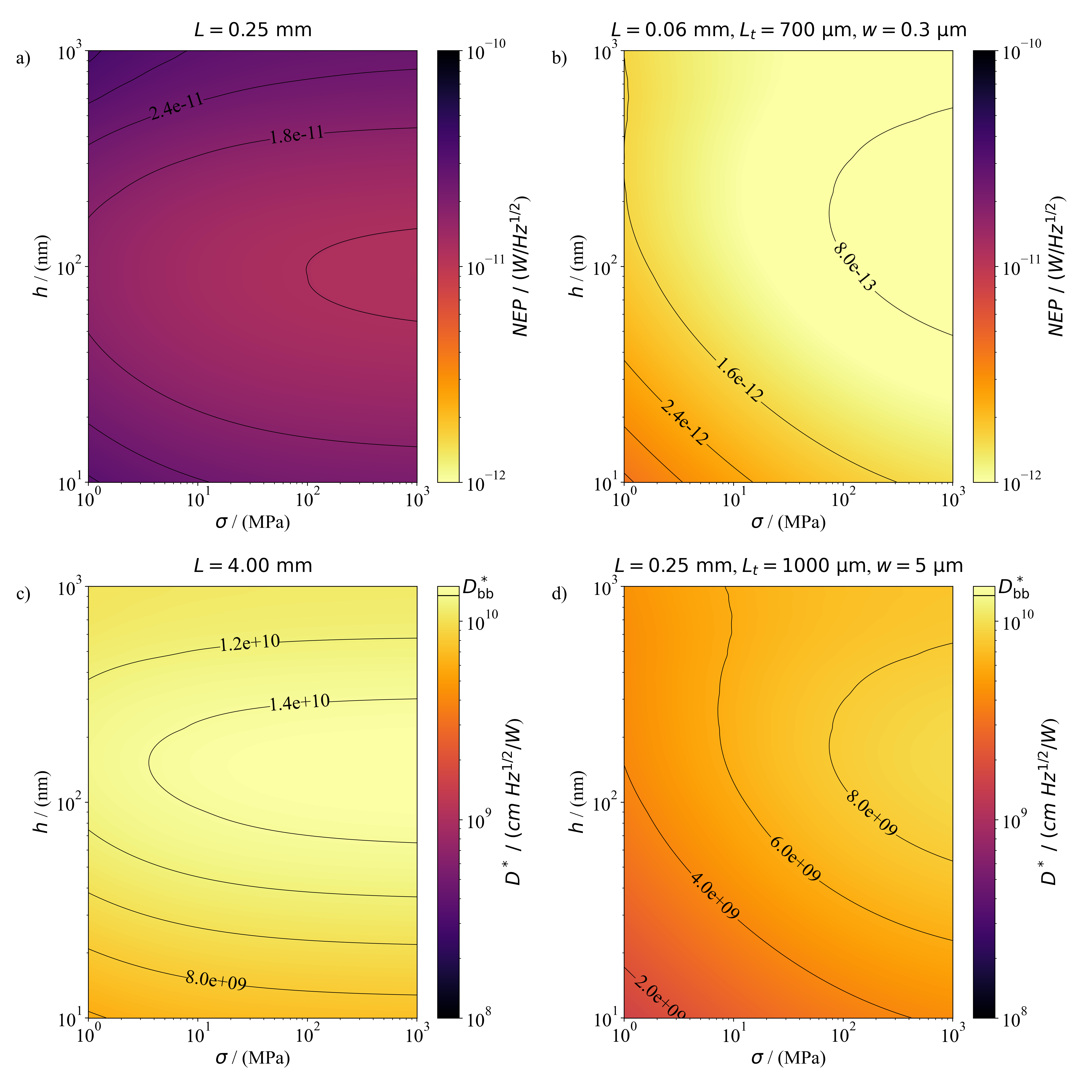}
    \caption{Heatmaps of NEP and specific detectivity $D^*$ for bare SiN detectors. Panels a) \& c) show NEP and $D^*$ for drumheads as functions of thickness $h$ and stress $\sigma$. Panels b)\& d) display the same for trampolines, with fixed, optimised values of tether length $L_t$ and width $w$. Colorbars for each figure of merit share the same scale across the drumhead and trampoline figures to allow direct comparison.}
    \label{fig:heatmaps_SiN}
\end{figure}
In this section, we consolidate the results from Sections \ref{sec:geometry} and \ref{sec:stress} to guide the reader through the various parameters investigated. We begin by focusing on NEP and $D^*$.

Figure \ref{fig:heatmaps_SiN} presents the results for bare SiN detectors. For drumheads, we show the figures of merit corresponding to the detection area side lengths $L$ that yielded the best performance: based on Figures \ref{fig:FOM_h}b and \ref{fig:FOM_sigma}b for NEP, and Figures \ref{fig:FOM_h}c and \ref{fig:FOM_sigma}c for $D^*$. The same selection criteria are applied to the trampolines, including the choice of tether length $L_t$ and width $w$ when varying thickness $h$ and prestress $\sigma$, and vice versa. Figure \ref{fig:heatmaps_SiN}a shows the NEP for drumheads with $L = \SI{0.25}{\milli\m}$. In line with the trends in Figures \ref{fig:FOM_h}b and \ref{fig:FOM_sigma}b, we observe a minimum NEP faround $h = \SI{100}{\nano\m}$ and for the highest stress value, $\sigma = \SI{1}{\giga\pascal}$. The corresponding specific detectivity is shown in Figure \ref{fig:heatmaps_SiN}c, and follows a similar trend. For this case, however, we used a structure with $L = \SI{4}{\milli\m}$, as suggested by the results in Figures \ref{fig:FOM_h}c and \ref{fig:FOM_sigma}c. 

The NEP of trampolines for varying $h$ and $\sigma$ is presented in Figure \ref{fig:heatmaps_SiN}b. Based on Figures \ref{fig:FOM_h}b and \ref{fig:FOM_sigma}b, we selected $L = \SI{0.062}{\milli\m}$, and from Figures \ref{fig:FOM_Lt}b the corresponding values of $L_T = \SI{700}{\micro\m}$. The tether width is $w = \SI{300}{\nano\m}$. The best performance is achieved for $\sigma = \SI{1}{\giga\pascal}$ and $h \simeq \SI{200}{\nano\m}$. Specific detectivity as a function of thickness and prestress is shown in Figure \ref{fig:heatmaps_SiN}d for a trampoline with $L = \SI{0.25}{\milli\m}$, $L_T = \SI{1000}{\micro\m}$, and $w = \SI{5}{\micro\m}$. It is important to note that, to enable direct comparison between drumhead and trampoline performance, the colorbars for NEP are set to the same range in Figures \ref{fig:heatmaps_SiN}a and \ref{fig:heatmaps_SiN}c. Comparing these plots, it is clear that trampolines generally exhibit better noise equivalent power than drumheads. Likewise, the colorbars for $D^*$ are matched across Figures \ref{fig:heatmaps_SiN}b and \ref{fig:heatmaps_SiN}d, where drumheads now clearly show superior performance.

\begin{figure}
    \centering
    \includegraphics[width=1\linewidth]{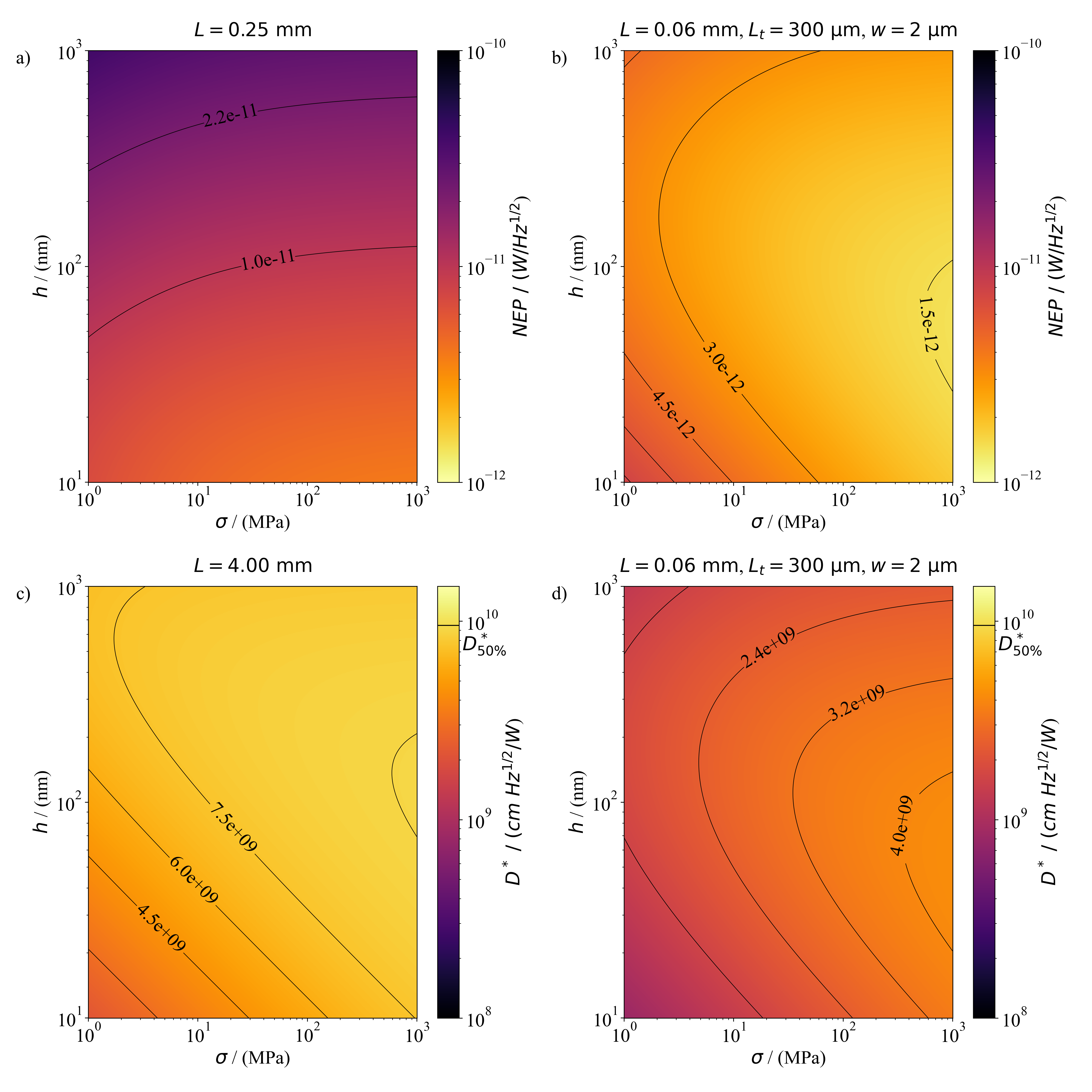}
    \caption{Heatmaps of NEP and specific detectivity $D^*$ for detectors coated with an ideal broadband absorber ($\alpha_{abs} = \epsilon_{abs} = 0.5$). Panels a) \& c) show drumhead performance as functions of $h$ and $\sigma$. Panels b) \& d) display the same for trampolines with optimised $L_t$ and $w$, as indicated in each plot title.}
    \label{fig:heatmaps_abs50}
\end{figure}
Next, we conduct the same analysis for the case of an ideal absorber, which gives the detectors a wavelength-independent emissivity and absorptance: $\epsilon_{abs} = \alpha_{abs} = 0.5$. All results are shown in Figure \ref{fig:heatmaps_abs50}. By comparing each plot here with its counterpart in Figure \ref{fig:heatmaps_SiN}, two main differences emerge. First, the overall performance of the detectors, both drumheads and trampolines, tends to degrade with the use of the absorber: the highest NEP values are higher, and the peak $D^*$ values are lower compared to the bare SiN case. This is due to the less favourable ratio between $\epsilon_{abs}$ and $\alpha_{abs}$ in the presence of the absorber. However, the use of an absorber remains advantageous when considering that these high absorptance values and the correspondingly good performance in both NEP and $D^*$ are maintained across a broad range of wavelengths. This makes the detector suitable for spectroscopy applications and significantly broadens its potential use cases. In contrast, the exceptional performance of bare SiN is confined to the narrow spectral region around its absorption peak. 

The second key difference is that, for trampolines, the optimal performance for both figures of merit is now achieved with smaller and thinner structures. For example, while the best specific detectivity in Figure \ref{fig:heatmaps_SiN}d was observed for $L = \SI{0.25}{\milli\m}$, the most favorable configuration in the absorber case is $L = \SI{0.062}{\milli\m}$. Likewise, the optimal values for tether length, thickness, and tether width shift to $L_T = \SI{300}{\micro\m}$, $h = \SI{10}{\nano\m}$, and $w = \SI{2}{\micro\m}$, respectively. This shift can again be attributed to the much higher emissivity of the detector: the enhanced radiative heat transfer moves the sweet spot, the equilibrium between heat conduction and radiation, toward smaller geometries.

\begin{figure}
    \centering
    \includegraphics[width=1\linewidth]{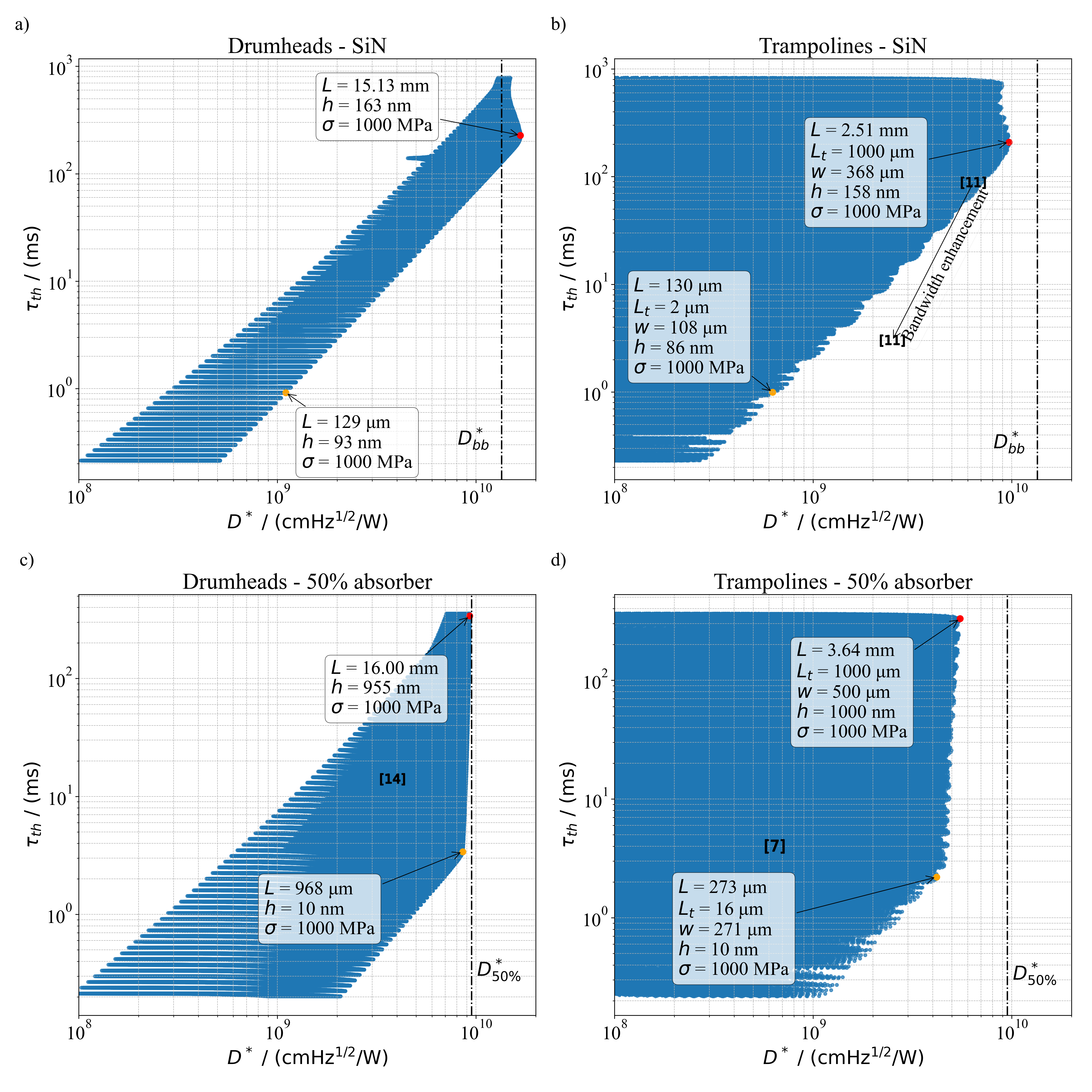}
    \caption{Pareto front plots for a) bare SiN drumheads; b) bare SiN trampolines; c) drumheads with absorber and d) trampolines with absorber, calculated for a logarithmic parameter sweep over $L = 62\,\mu\mathrm{m}\text{–}16\,\mathrm{mm}$, 
$h = 10\,\mathrm{nm}\text{–}1\,\mu\mathrm{m}$, 
$\sigma = 1\,\mathrm{MPa}\text{–}1\,\mathrm{GPa}$, 
$w = 5\,\mu\mathrm{m}\text{–}500\,\mu\mathrm{m}$, 
and $L_t = 1\,\mu\mathrm{m}\text{–}1\,\mathrm{mm}$. In each plot, the dotted, vertical line represents the room temperature detection limit (two-sides coupled). In red, we have highlighted the absolute maxima, while in yellow, we have highlighted the relevant local maxima. The combination of variables of both absolute and local maxima is reported in the corresponding annotation box. References indicate relevant state-of-the-art results, placed at the positions in the plot matching their reported values.}
    \label{fig:pareto}
\end{figure}

Finally, we evaluated the sensitivity $D^*$ and thermal response time $\tau_{th}$ for combinations of geometrical and material parameters for both drumhead and trampoline designs (bare SiN and absorber-integrated), as summarized in Figure \ref{fig:pareto}. Depending on the application, one may prioritize either maximal sensitivity or minimal response time. Example parameter sets optimized purely for sensitivity, irrespective of speed, are indicated by red points, whereas designs that trade sensitivity for faster response are shown in yellow. To ensure that all simulated geometries fall within realistic fabrication limits, the explored parameter space was chosen to lie well inside the bounds established by the most extreme demonstrated SiN membrane technologies, which span centimeter-scale suspended LPCVD SiN membranes (up to $6\times 6$~cm at 200~nm thickness) \cite{moura2018centimeter,norder2025pentagonal} down to ultrathin films with thicknesses as low as 5~nm \cite{dutt2023ultrathin,tong2004silicon}.

Figure \ref{fig:pareto}a shows the results for a selection of the variable combinations for bare SiN drumheads. We can readily observe that here we obtain the absolute highest value of $D^*$, highlighted in red in the plot, surpassing the fundamental room temperature limit of an ideal black body, $D^*_{bb}$, marked as the vertical dotted black line. The particular combination of variables for which a SiN drumhead can reach this performance is annotated in the relative text box. This structure has a relatively slow thermal time constant, being $\tau_{th}>\SI{100}{\milli\s}$. We therefore also highlighted in yellow a local maximum for $\tau_{th}<\SI{1}{\milli\s}$. This detector is more two orders of magnitude faster than the other highlighted one, while still having a high detectivity $D^*>\SI{1e9}{\centi\m\sqrt\hertz\per\watt}$. 
The Pareto front for bare SiN trampolines is shown in Figure \ref{fig:pareto}b. Similarly to SiN drumheads, the highest value of $D^*$ is obtained for a structure (red dot) that is slow, $\tau_{th}>\SI{100}{\milli\s}$. In order to obtain a detector with $\tau_{th}<\SI{1}{\milli\s}$ (yellow dot) we have to sacrifice the specific detectivity and accept having $D^*<\SI{1e9}{\centi\m\sqrt\hertz\per\watt}$.

In Figure \ref{fig:pareto}c, the Pareto front plot for drumheads covered with an absorber is displayed. The vertical dotted line marks the room temperature sensitivity limit for a detector with a 50\% absorber, $D^*_{50\%}$.
We notice that this plot's front approaches the theoretical limit over a large range of time constants. This allows the design of an optimal drumhead detector with a time constant as low as 3~ms and that operates close to the theoretical detectivity limit (yellow dot) in contrast to the slow design with the globally maximum $D^*$-value.

In Figure \ref{fig:pareto}d, we show the results for trampolines with an absorber. Here too, the absolute maximum (red dot) is obtained for a structure that has a high thermal time constant. 
In this case as well, we highlight with a yellow dot a specific structure with standard fabrication geometry and stress values that is two orders of magnitude faster than the detector with the highest $D^*$, while exhibiting nearly the same detectivity.

In Figures~\ref{fig:pareto}b, c, \& d, we display references to relevant state-of-the-art results, which are summarised in Table~\ref{tab:pareto} and compared to the predicted theoretical values by our model. 
The full geometrical parameters are listed in the SI. 
Of particular relevance is the recent work by Zhang \textit{et al.} \cite{zhang2025enhanced}, who demonstrated that the bandwidth limitation imposed by the thermal time constant can be overcome when operating the resonator in the temperature fluctuation-limited regime. Using SiN trampoline resonators, the authors achieved a specific detectivity located at the Pareto front. Through a bandwidth enhancement by a factor of thirty, they achieved a performance beyond the front. This is illustrated in Figure \ref{fig:pareto}b, where the two references are connected by a black arrow, representing the standard and enhanced bandwidth measurements. In the case of the drumhead example,\cite{martini2025uncooled} the deviation between theory and experiment arises mainly because the detectors were not operated at the fundamental noise limit; additional transduction-related noise sources slightly degrade the measured performance. Also for trampolines,\cite{piller2022thermal} the difference between reported and predicted values is primarily due to transduction noise, which prevents operation at the fundamental limit. Furthermore, the presence of ($\SI{100}{\nano\m}$) thick Au electrodes on the tethers increases thermal conduction, yielding a faster thermal response than that predicted when only the metal absorber is considered. 

All highlighted resonator designs in Figure~\ref{fig:pareto} (yellow and red points) are made of high-stress SiN. As shown in Figure~\ref{fig:ADs for sigma}, this results in low frequency noise, which requires that no other noise source, such as transduction noise, is dominating. Using low-noise optical interferometry for readout, the optimal performance of the specific nanomechanical IR detectors highlighted in Figure~\ref{fig:pareto} should be achievable.

\begin{table}
    \centering
    \begin{tabular}{c|c|c||c|c}
        Detector
            & \multicolumn{2}{c||}{Experimental}
            & \multicolumn{2}{c}{Theory} \\
            & $D^*~[cm\sqrt{Hz}/W]$
            & $\tau_{th}~[ms]$
            & $D^*~[cm\sqrt{Hz}/W]$
            & $\tau_{th}~[ms]$ \\
        \hline\hline

        SiN trampolines\cite{zhang2025enhanced}
            & $6.4\times10^{9}$
            & $88$
            & $5.4\times10^{9}$
            & $97$ \\
        \hline

        Drumheads w/ absorber\cite{martini2025uncooled}
            & $3.8\times10^{9}$
            & $14$
            & $7.6\times10^{9}$
            & $14$ \\
        \hline

        Trampolines w/ absorber\cite{piller2022thermal}
            & $6.4\times10^{8}$
            & $4$
            & $1.9\times10^{9}$
            & $17$ \\
    \end{tabular}

    \caption{Comparison between experimentally measured and analytically predicted specific detectivity and thermal time constant for the detectors reported in Fig.~\ref{fig:pareto}. The geometrical and stress parameters for each structure are reported in the SI. The experimental value reported in \cite{zhang2025enhanced} was rescaled considering only the central detection area.}
    \label{tab:pareto}
\end{table}

\section{Conclusions}
\label{conclusion}

In conclusion, we have shown that the current state-of-the-art in thermomechanical IR detectors has not yet reached its full potential. We have highlighted the architectural pathway to achieving the fundamental room-temperature detection limit and identified several aspects where researchers can focus to improve existing designs. For both bare SiN and devices with a broadband absorber, drumhead resonators outperform their trampoline counterparts, exhibiting higher specific detectivity and shorter thermal time constants. Beyond clarifying the basic differences between drumhead and trampoline resonators, we have underscored the fundamental trade-off between achieving peak performance over a narrow range of detectable wavelengths and sacrificing a small portion of detectable power in favour of a significantly broader operational bandwidth.  Crucially, this work bridges a long-standing gap in the literature by combining analytical modelling with practical design insights, laying the groundwork for the development of detectors that approach the fundamental limits of thermal sensitivity. We anticipate that this guide will accelerate innovation in nanomechanical sensing and foster the deployment of next-generation IR detection technologies in a broad range of scientific and technological domains.

\section{Supplementary Material}
In the Supplementary Material, we provide additional details and figures on several aspects of this work, including the measurement of the Duffing parameter for trampolines, the influence of tether width on the figures of merit, the calculation of absorptance and emissivity for different SiN thicknesses, the noise affecting the investigated resonators, the dependence of power responsivity on various parameters, and additional results for $\tau_{th}$ and $R_P$ in the case of a detector with a metal absorber.

\section{acknowledgments}
The authors would like to thank Dr. Hajrudin Besic, Dr. Robert G. West, and Jelena Timarac Popovic for the many fruitful discussions and valuable support during the writing of this work. This work received funding from the Novo Nordisk Foundation under the project MASMONADE with
project number NNF22OC0077964. 

\section{Authors' declarations}
\subsection{Conflict of Interest}
The authors have no conflicts to disclose.
\subsection{Author Contributions}
P.M. contributed with measurements, production of the results and figures, and the overall structure and writing of the paper; K.K. contributed with measurements, data analysis and assisted with the writing; S.S. conceived the idea of the paper, discussed the results, supervised the work and co-wrote the paper.
\subsection{Data availability}
All the data that supports the findings of this study are available from the corresponding author upon reasonable request.

\bibliography{bib}

\begin{thebibliography}{46}%
\makeatletter
\providecommand \@ifxundefined [1]{%
 \@ifx{#1\undefined}
}%
\providecommand \@ifnum [1]{%
 \ifnum #1\expandafter \@firstoftwo
 \else \expandafter \@secondoftwo
 \fi
}%
\providecommand \@ifx [1]{%
 \ifx #1\expandafter \@firstoftwo
 \else \expandafter \@secondoftwo
 \fi
}%
\providecommand \natexlab [1]{#1}%
\providecommand \enquote  [1]{``#1''}%
\providecommand \bibnamefont  [1]{#1}%
\providecommand \bibfnamefont [1]{#1}%
\providecommand \citenamefont [1]{#1}%
\providecommand \href@noop [0]{\@secondoftwo}%
\providecommand \href [0]{\begingroup \@sanitize@url \@href}%
\providecommand \@href[1]{\@@startlink{#1}\@@href}%
\providecommand \@@href[1]{\endgroup#1\@@endlink}%
\providecommand \@sanitize@url [0]{\catcode `\\12\catcode `\$12\catcode `\&12\catcode `\#12\catcode `\^12\catcode `\_12\catcode `\%12\relax}%
\providecommand \@@startlink[1]{}%
\providecommand \@@endlink[0]{}%
\providecommand \url  [0]{\begingroup\@sanitize@url \@url }%
\providecommand \@url [1]{\endgroup\@href {#1}{\urlprefix }}%
\providecommand \urlprefix  [0]{URL }%
\providecommand \Eprint [0]{\href }%
\providecommand \doibase [0]{http://dx.doi.org/}%
\providecommand \selectlanguage [0]{\@gobble}%
\providecommand \bibinfo  [0]{\@secondoftwo}%
\providecommand \bibfield  [0]{\@secondoftwo}%
\providecommand \translation [1]{[#1]}%
\providecommand \BibitemOpen [0]{}%
\providecommand \bibitemStop [0]{}%
\providecommand \bibitemNoStop [0]{.\EOS\space}%
\providecommand \EOS [0]{\spacefactor3000\relax}%
\providecommand \BibitemShut  [1]{\csname bibitem#1\endcsname}%
\let\auto@bib@innerbib\@empty
\bibitem [{\citenamefont {Kruse}\ and\ \citenamefont {Skatrud}(1997)}]{kruse1997uncooled}%
  \BibitemOpen
  \bibfield  {author} {\bibinfo {author} {\bibfnamefont {P.~W.}\ \bibnamefont {Kruse}}\ and\ \bibinfo {author} {\bibfnamefont {D.~D.}\ \bibnamefont {Skatrud}},\ }\href@noop {} {\emph {\bibinfo {title} {Uncooled infrared imaging arrays and systems}}},\ Vol.~\bibinfo {volume} {47}\ (\bibinfo  {publisher} {Academic press},\ \bibinfo {year} {1997})\BibitemShut {NoStop}%
\bibitem [{\citenamefont {Rogalski}(2019)}]{rogalski2019infrared}%
  \BibitemOpen
  \bibfield  {author} {\bibinfo {author} {\bibfnamefont {A.}~\bibnamefont {Rogalski}},\ }\href@noop {} {\emph {\bibinfo {title} {Infrared and terahertz detectors}}}\ (\bibinfo  {publisher} {CRC press},\ \bibinfo {year} {2019})\BibitemShut {NoStop}%
\bibitem [{\citenamefont {Vig}, \citenamefont {Filler},\ and\ \citenamefont {Kim}(1996)}]{vig1996uncooled}%
  \BibitemOpen
  \bibfield  {author} {\bibinfo {author} {\bibfnamefont {J.~R.}\ \bibnamefont {Vig}}, \bibinfo {author} {\bibfnamefont {R.}~\bibnamefont {Filler}}, \ and\ \bibinfo {author} {\bibfnamefont {Y.}~\bibnamefont {Kim}},\ }\bibfield  {title} {\enquote {\bibinfo {title} {Uncooled ir imaging array based on quartz microresonators},}\ }\href@noop {} {\bibfield  {journal} {\bibinfo  {journal} {Journal of Microelectromechanical Systems}\ }\textbf {\bibinfo {volume} {5}},\ \bibinfo {pages} {131--137} (\bibinfo {year} {1996})}\BibitemShut {NoStop}%
\bibitem [{\citenamefont {Zhang}\ \emph {et~al.}(2013)\citenamefont {Zhang}, \citenamefont {Myers}, \citenamefont {Sader},\ and\ \citenamefont {Roukes}}]{zhang2013nanomechanical}%
  \BibitemOpen
  \bibfield  {author} {\bibinfo {author} {\bibfnamefont {X.}~\bibnamefont {Zhang}}, \bibinfo {author} {\bibfnamefont {E.}~\bibnamefont {Myers}}, \bibinfo {author} {\bibfnamefont {J.}~\bibnamefont {Sader}}, \ and\ \bibinfo {author} {\bibfnamefont {M.}~\bibnamefont {Roukes}},\ }\bibfield  {title} {\enquote {\bibinfo {title} {Nanomechanical torsional resonators for frequency-shift infrared thermal sensing},}\ }\href@noop {} {\bibfield  {journal} {\bibinfo  {journal} {Nano letters}\ }\textbf {\bibinfo {volume} {13}},\ \bibinfo {pages} {1528--1534} (\bibinfo {year} {2013})}\BibitemShut {NoStop}%
\bibitem [{\citenamefont {Hui}\ and\ \citenamefont {Rinaldi}(2013)}]{hui2013high}%
  \BibitemOpen
  \bibfield  {author} {\bibinfo {author} {\bibfnamefont {Y.}~\bibnamefont {Hui}}\ and\ \bibinfo {author} {\bibfnamefont {M.}~\bibnamefont {Rinaldi}},\ }\bibfield  {title} {\enquote {\bibinfo {title} {High performance nems resonant infrared detector based on an aluminum nitride nano-plate resonator},}\ }in\ \href@noop {} {\emph {\bibinfo {booktitle} {2013 Transducers \& Eurosensors XXVII: The 17th International Conference on Solid-State Sensors, Actuators and Microsystems (TRANSDUCERS \& EUROSENSORS XXVII)}}}\ (\bibinfo {organization} {IEEE},\ \bibinfo {year} {2013})\ pp.\ \bibinfo {pages} {968--971}\BibitemShut {NoStop}%
\bibitem [{\citenamefont {Vicarelli}, \citenamefont {Tredicucci},\ and\ \citenamefont {Pitanti}(2022)}]{vicarelli2022micromechanical}%
  \BibitemOpen
  \bibfield  {author} {\bibinfo {author} {\bibfnamefont {L.}~\bibnamefont {Vicarelli}}, \bibinfo {author} {\bibfnamefont {A.}~\bibnamefont {Tredicucci}}, \ and\ \bibinfo {author} {\bibfnamefont {A.}~\bibnamefont {Pitanti}},\ }\bibfield  {title} {\enquote {\bibinfo {title} {Micromechanical bolometers for subterahertz detection at room temperature},}\ }\href@noop {} {\bibfield  {journal} {\bibinfo  {journal} {ACS photonics}\ }\textbf {\bibinfo {volume} {9}},\ \bibinfo {pages} {360--367} (\bibinfo {year} {2022})}\BibitemShut {NoStop}%
\bibitem [{\citenamefont {Piller}\ \emph {et~al.}(2022)\citenamefont {Piller}, \citenamefont {Hiesberger}, \citenamefont {Wistrela}, \citenamefont {Martini}, \citenamefont {Luhmann},\ and\ \citenamefont {Schmid}}]{piller2022thermal}%
  \BibitemOpen
  \bibfield  {author} {\bibinfo {author} {\bibfnamefont {M.}~\bibnamefont {Piller}}, \bibinfo {author} {\bibfnamefont {J.}~\bibnamefont {Hiesberger}}, \bibinfo {author} {\bibfnamefont {E.}~\bibnamefont {Wistrela}}, \bibinfo {author} {\bibfnamefont {P.}~\bibnamefont {Martini}}, \bibinfo {author} {\bibfnamefont {N.}~\bibnamefont {Luhmann}}, \ and\ \bibinfo {author} {\bibfnamefont {S.}~\bibnamefont {Schmid}},\ }\bibfield  {title} {\enquote {\bibinfo {title} {Thermal ir detection with nanoelectromechanical silicon nitride trampoline resonators},}\ }\href@noop {} {\bibfield  {journal} {\bibinfo  {journal} {IEEE Sensors Journal}\ }\textbf {\bibinfo {volume} {23}},\ \bibinfo {pages} {1066--1071} (\bibinfo {year} {2022})}\BibitemShut {NoStop}%
\bibitem [{\citenamefont {Li}, \citenamefont {Zhang},\ and\ \citenamefont {Hirakawa}(2023)}]{li2023terahertz}%
  \BibitemOpen
  \bibfield  {author} {\bibinfo {author} {\bibfnamefont {C.}~\bibnamefont {Li}}, \bibinfo {author} {\bibfnamefont {Y.}~\bibnamefont {Zhang}}, \ and\ \bibinfo {author} {\bibfnamefont {K.}~\bibnamefont {Hirakawa}},\ }\bibfield  {title} {\enquote {\bibinfo {title} {Terahertz detectors using microelectromechanical system resonators},}\ }\href@noop {} {\bibfield  {journal} {\bibinfo  {journal} {Sensors}\ }\textbf {\bibinfo {volume} {23}},\ \bibinfo {pages} {5938} (\bibinfo {year} {2023})}\BibitemShut {NoStop}%
\bibitem [{\citenamefont {Zhang}\ \emph {et~al.}(2024)\citenamefont {Zhang}, \citenamefont {Yalavarthi}, \citenamefont {Giroux}, \citenamefont {Cui}, \citenamefont {Stephan}, \citenamefont {Maleki}, \citenamefont {Weck}, \citenamefont {M{\'e}nard},\ and\ \citenamefont {St-Gelais}}]{zhang2024high}%
  \BibitemOpen
  \bibfield  {author} {\bibinfo {author} {\bibfnamefont {C.}~\bibnamefont {Zhang}}, \bibinfo {author} {\bibfnamefont {E.~K.}\ \bibnamefont {Yalavarthi}}, \bibinfo {author} {\bibfnamefont {M.}~\bibnamefont {Giroux}}, \bibinfo {author} {\bibfnamefont {W.}~\bibnamefont {Cui}}, \bibinfo {author} {\bibfnamefont {M.}~\bibnamefont {Stephan}}, \bibinfo {author} {\bibfnamefont {A.}~\bibnamefont {Maleki}}, \bibinfo {author} {\bibfnamefont {A.}~\bibnamefont {Weck}}, \bibinfo {author} {\bibfnamefont {J.-M.}\ \bibnamefont {M{\'e}nard}}, \ and\ \bibinfo {author} {\bibfnamefont {R.}~\bibnamefont {St-Gelais}},\ }\bibfield  {title} {\enquote {\bibinfo {title} {High detectivity terahertz radiation sensing using frequency-noise-optimized nanomechanical resonators},}\ }\href@noop {} {\bibfield  {journal} {\bibinfo  {journal} {APL Photonics}\ }\textbf {\bibinfo {volume} {9}} (\bibinfo {year} {2024})}\BibitemShut {NoStop}%
\bibitem [{\citenamefont {Das}\ \emph {et~al.}(2023)\citenamefont {Das}, \citenamefont {Mah}, \citenamefont {Hunt},\ and\ \citenamefont {Talghader}}]{das2023thermodynamically}%
  \BibitemOpen
  \bibfield  {author} {\bibinfo {author} {\bibfnamefont {A.}~\bibnamefont {Das}}, \bibinfo {author} {\bibfnamefont {M.~L.}\ \bibnamefont {Mah}}, \bibinfo {author} {\bibfnamefont {J.}~\bibnamefont {Hunt}}, \ and\ \bibinfo {author} {\bibfnamefont {J.~J.}\ \bibnamefont {Talghader}},\ }\bibfield  {title} {\enquote {\bibinfo {title} {Thermodynamically limited uncooled infrared detector using an ultra-low mass perforated subwavelength absorber},}\ }\href@noop {} {\bibfield  {journal} {\bibinfo  {journal} {Optica}\ }\textbf {\bibinfo {volume} {10}},\ \bibinfo {pages} {1018--1028} (\bibinfo {year} {2023})}\BibitemShut {NoStop}%
\bibitem [{\citenamefont {Zhang}\ \emph {et~al.}(2025)\citenamefont {Zhang}, \citenamefont {Louis-Seize}, \citenamefont {Saleh}, \citenamefont {Brazeau}, \citenamefont {Hodges}, \citenamefont {Turgeon-Roy},\ and\ \citenamefont {St-Gelais}}]{zhang2025enhanced}%
  \BibitemOpen
  \bibfield  {author} {\bibinfo {author} {\bibfnamefont {C.}~\bibnamefont {Zhang}}, \bibinfo {author} {\bibfnamefont {Z.}~\bibnamefont {Louis-Seize}}, \bibinfo {author} {\bibfnamefont {Y.}~\bibnamefont {Saleh}}, \bibinfo {author} {\bibfnamefont {M.}~\bibnamefont {Brazeau}}, \bibinfo {author} {\bibfnamefont {T.}~\bibnamefont {Hodges}}, \bibinfo {author} {\bibfnamefont {M.}~\bibnamefont {Turgeon-Roy}}, \ and\ \bibinfo {author} {\bibfnamefont {R.}~\bibnamefont {St-Gelais}},\ }\bibfield  {title} {\enquote {\bibinfo {title} {Enhanced bandwidth in radiation sensors operating at the fundamental temperature fluctuation noise limit},}\ }\href@noop {} {\bibfield  {journal} {\bibinfo  {journal} {Nano Letters}\ } (\bibinfo {year} {2025})}\BibitemShut {NoStop}%
\bibitem [{\citenamefont {Beliaev}\ \emph {et~al.}(2022)\citenamefont {Beliaev}, \citenamefont {Shkondin}, \citenamefont {Lavrinenko},\ and\ \citenamefont {Takayama}}]{beliaev2022optical}%
  \BibitemOpen
  \bibfield  {author} {\bibinfo {author} {\bibfnamefont {L.~Y.}\ \bibnamefont {Beliaev}}, \bibinfo {author} {\bibfnamefont {E.}~\bibnamefont {Shkondin}}, \bibinfo {author} {\bibfnamefont {A.~V.}\ \bibnamefont {Lavrinenko}}, \ and\ \bibinfo {author} {\bibfnamefont {O.}~\bibnamefont {Takayama}},\ }\bibfield  {title} {\enquote {\bibinfo {title} {Optical, structural and composition properties of silicon nitride films deposited by reactive radio-frequency sputtering, low pressure and plasma-enhanced chemical vapor deposition},}\ }\href@noop {} {\bibfield  {journal} {\bibinfo  {journal} {Thin Solid Films}\ }\textbf {\bibinfo {volume} {763}},\ \bibinfo {pages} {139568} (\bibinfo {year} {2022})}\BibitemShut {NoStop}%
\bibitem [{\citenamefont {Kanellopulos}\ \emph {et~al.}(2025)\citenamefont {Kanellopulos}, \citenamefont {Ladinig}, \citenamefont {Emminger}, \citenamefont {Martini}, \citenamefont {West},\ and\ \citenamefont {Schmid}}]{kanellopulos2025comparative}%
  \BibitemOpen
  \bibfield  {author} {\bibinfo {author} {\bibfnamefont {K.}~\bibnamefont {Kanellopulos}}, \bibinfo {author} {\bibfnamefont {F.}~\bibnamefont {Ladinig}}, \bibinfo {author} {\bibfnamefont {S.}~\bibnamefont {Emminger}}, \bibinfo {author} {\bibfnamefont {P.}~\bibnamefont {Martini}}, \bibinfo {author} {\bibfnamefont {R.~G.}\ \bibnamefont {West}}, \ and\ \bibinfo {author} {\bibfnamefont {S.}~\bibnamefont {Schmid}},\ }\bibfield  {title} {\enquote {\bibinfo {title} {Comparative analysis of nanomechanical resonators: sensitivity, response time, and practical considerations in photothermal sensing},}\ }\href@noop {} {\bibfield  {journal} {\bibinfo  {journal} {Microsystems \& Nanoengineering}\ }\textbf {\bibinfo {volume} {11}},\ \bibinfo {pages} {28} (\bibinfo {year} {2025})}\BibitemShut {NoStop}%
\bibitem [{\citenamefont {Martini}\ \emph {et~al.}(2025)\citenamefont {Martini}, \citenamefont {Kanellopulos}, \citenamefont {Emminger}, \citenamefont {Luhmann}, \citenamefont {Piller}, \citenamefont {West},\ and\ \citenamefont {Schmid}}]{martini2025uncooled}%
  \BibitemOpen
  \bibfield  {author} {\bibinfo {author} {\bibfnamefont {P.}~\bibnamefont {Martini}}, \bibinfo {author} {\bibfnamefont {K.}~\bibnamefont {Kanellopulos}}, \bibinfo {author} {\bibfnamefont {S.}~\bibnamefont {Emminger}}, \bibinfo {author} {\bibfnamefont {N.}~\bibnamefont {Luhmann}}, \bibinfo {author} {\bibfnamefont {M.}~\bibnamefont {Piller}}, \bibinfo {author} {\bibfnamefont {R.~G.}\ \bibnamefont {West}}, \ and\ \bibinfo {author} {\bibfnamefont {S.}~\bibnamefont {Schmid}},\ }\bibfield  {title} {\enquote {\bibinfo {title} {Uncooled thermal infrared detection near the fundamental limit using a silicon nitride nanomechanical resonator with a broadband absorber},}\ }\href@noop {} {\bibfield  {journal} {\bibinfo  {journal} {Communications Physics}\ }\textbf {\bibinfo {volume} {8}},\ \bibinfo {pages} {166} (\bibinfo {year} {2025})}\BibitemShut {NoStop}%
\bibitem [{\citenamefont {Datskos}\ and\ \citenamefont {Lavrik}(2003)}]{datskos2003detectors}%
  \BibitemOpen
  \bibfield  {author} {\bibinfo {author} {\bibfnamefont {P.~G.}\ \bibnamefont {Datskos}}\ and\ \bibinfo {author} {\bibfnamefont {N.~V.}\ \bibnamefont {Lavrik}},\ }\bibfield  {title} {\enquote {\bibinfo {title} {Detectors—figures of merit},}\ }\href@noop {} {\bibfield  {journal} {\bibinfo  {journal} {Encyclopedia of Optical Engineering}\ }\textbf {\bibinfo {volume} {349}} (\bibinfo {year} {2003})}\BibitemShut {NoStop}%
\bibitem [{\citenamefont {Kruse}(2004)}]{kruse2004can}%
  \BibitemOpen
  \bibfield  {author} {\bibinfo {author} {\bibfnamefont {P.~W.}\ \bibnamefont {Kruse}},\ }\bibfield  {title} {\enquote {\bibinfo {title} {Can the 300-k radiating background noise limit be attained by uncooled thermal imagers?}}\ }in\ \href@noop {} {\emph {\bibinfo {booktitle} {Infrared Technology and Applications XXX}}},\ Vol.\ \bibinfo {volume} {5406}\ (\bibinfo {organization} {SPIE},\ \bibinfo {year} {2004})\ pp.\ \bibinfo {pages} {437--446}\BibitemShut {NoStop}%
\bibitem [{\citenamefont {Skidmore}\ \emph {et~al.}(2003)\citenamefont {Skidmore}, \citenamefont {Gildemeister}, \citenamefont {Lee}, \citenamefont {Myers},\ and\ \citenamefont {Richards}}]{skidmore2003superconducting}%
  \BibitemOpen
  \bibfield  {author} {\bibinfo {author} {\bibfnamefont {J.}~\bibnamefont {Skidmore}}, \bibinfo {author} {\bibfnamefont {J.}~\bibnamefont {Gildemeister}}, \bibinfo {author} {\bibfnamefont {A.}~\bibnamefont {Lee}}, \bibinfo {author} {\bibfnamefont {M.}~\bibnamefont {Myers}}, \ and\ \bibinfo {author} {\bibfnamefont {P.}~\bibnamefont {Richards}},\ }\bibfield  {title} {\enquote {\bibinfo {title} {Superconducting bolometer for far-infrared fourier transform spectroscopy},}\ }\href@noop {} {\bibfield  {journal} {\bibinfo  {journal} {Applied physics letters}\ }\textbf {\bibinfo {volume} {82}},\ \bibinfo {pages} {469--471} (\bibinfo {year} {2003})}\BibitemShut {NoStop}%
\bibitem [{\citenamefont {Schmid}, \citenamefont {Villanueva},\ and\ \citenamefont {Roukes}(2023)}]{schmid2023fundamentals}%
  \BibitemOpen
  \bibfield  {author} {\bibinfo {author} {\bibfnamefont {S.}~\bibnamefont {Schmid}}, \bibinfo {author} {\bibfnamefont {L.~G.}\ \bibnamefont {Villanueva}}, \ and\ \bibinfo {author} {\bibfnamefont {M.~L.}\ \bibnamefont {Roukes}},\ }\href@noop {} {\enquote {\bibinfo {title} {Fundamentals of nanomechanical resonators},}\ } (\bibinfo {year} {2023})\BibitemShut {NoStop}%
\bibitem [{\citenamefont {Kanellopulos}(2025)}]{Kanellopulos2025dissertation}%
  \BibitemOpen
  \bibfield  {author} {\bibinfo {author} {\bibfnamefont {K.}~\bibnamefont {Kanellopulos}},\ }\href@noop {} {\emph {\bibinfo {title} {Nanomechanical Photothermal Sensing}}}\ (\bibinfo  {publisher} {Technical University of Vienna},\ \bibinfo {year} {2025})\ \bibinfo {note} {ph.D. Dissertation}\BibitemShut {NoStop}%
\bibitem [{\citenamefont {Martini}, \citenamefont {Kanellopulos},\ and\ \citenamefont {Schmid}(2023)}]{martini2023towards}%
  \BibitemOpen
  \bibfield  {author} {\bibinfo {author} {\bibfnamefont {P.}~\bibnamefont {Martini}}, \bibinfo {author} {\bibfnamefont {K.}~\bibnamefont {Kanellopulos}}, \ and\ \bibinfo {author} {\bibfnamefont {S.}~\bibnamefont {Schmid}},\ }\bibfield  {title} {\enquote {\bibinfo {title} {Towards photon-noise limited thermal ir detection with optomechanical resonators},}\ }in\ \href@noop {} {\emph {\bibinfo {booktitle} {2023 IEEE SENSORS}}}\ (\bibinfo {organization} {IEEE},\ \bibinfo {year} {2023})\ pp.\ \bibinfo {pages} {1--4}\BibitemShut {NoStop}%
\bibitem [{\citenamefont {Be{\v{s}}i{\'c}}\ \emph {et~al.}(2023)\citenamefont {Be{\v{s}}i{\'c}}, \citenamefont {Demir}, \citenamefont {Steurer}, \citenamefont {Luhmann},\ and\ \citenamefont {Schmid}}]{bevsic2023schemes}%
  \BibitemOpen
  \bibfield  {author} {\bibinfo {author} {\bibfnamefont {H.}~\bibnamefont {Be{\v{s}}i{\'c}}}, \bibinfo {author} {\bibfnamefont {A.}~\bibnamefont {Demir}}, \bibinfo {author} {\bibfnamefont {J.}~\bibnamefont {Steurer}}, \bibinfo {author} {\bibfnamefont {N.}~\bibnamefont {Luhmann}}, \ and\ \bibinfo {author} {\bibfnamefont {S.}~\bibnamefont {Schmid}},\ }\bibfield  {title} {\enquote {\bibinfo {title} {Schemes for tracking resonance frequency for micro-and nanomechanical resonators},}\ }\href@noop {} {\bibfield  {journal} {\bibinfo  {journal} {Physical Review Applied}\ }\textbf {\bibinfo {volume} {20}},\ \bibinfo {pages} {024023} (\bibinfo {year} {2023})}\BibitemShut {NoStop}%
\bibitem [{\citenamefont {Demir}(2021)}]{demir2021understanding}%
  \BibitemOpen
  \bibfield  {author} {\bibinfo {author} {\bibfnamefont {A.}~\bibnamefont {Demir}},\ }\bibfield  {title} {\enquote {\bibinfo {title} {Understanding fundamental trade-offs in nanomechanical resonant sensors},}\ }\href@noop {} {\bibfield  {journal} {\bibinfo  {journal} {Journal of Applied Physics}\ }\textbf {\bibinfo {volume} {129}} (\bibinfo {year} {2021})}\BibitemShut {NoStop}%
\bibitem [{\citenamefont {Lu}\ \emph {et~al.}(2020)\citenamefont {Lu}, \citenamefont {Shao}, \citenamefont {Amabili}, \citenamefont {Yue},\ and\ \citenamefont {Guo}}]{lu2020nonlinear}%
  \BibitemOpen
  \bibfield  {author} {\bibinfo {author} {\bibfnamefont {Y.}~\bibnamefont {Lu}}, \bibinfo {author} {\bibfnamefont {Q.}~\bibnamefont {Shao}}, \bibinfo {author} {\bibfnamefont {M.}~\bibnamefont {Amabili}}, \bibinfo {author} {\bibfnamefont {H.}~\bibnamefont {Yue}}, \ and\ \bibinfo {author} {\bibfnamefont {H.}~\bibnamefont {Guo}},\ }\bibfield  {title} {\enquote {\bibinfo {title} {Nonlinear vibration control effects of membrane structures with in-plane pvdf actuators: A parametric study},}\ }\href@noop {} {\bibfield  {journal} {\bibinfo  {journal} {International Journal of Non-Linear Mechanics}\ }\textbf {\bibinfo {volume} {122}},\ \bibinfo {pages} {103466} (\bibinfo {year} {2020})}\BibitemShut {NoStop}%
\bibitem [{\citenamefont {Manzaneque}\ \emph {et~al.}(2023)\citenamefont {Manzaneque}, \citenamefont {Ghatkesar}, \citenamefont {Alijani}, \citenamefont {Xu}, \citenamefont {Norte},\ and\ \citenamefont {Steeneken}}]{manzaneque2023resolution}%
  \BibitemOpen
  \bibfield  {author} {\bibinfo {author} {\bibfnamefont {T.}~\bibnamefont {Manzaneque}}, \bibinfo {author} {\bibfnamefont {M.~K.}\ \bibnamefont {Ghatkesar}}, \bibinfo {author} {\bibfnamefont {F.}~\bibnamefont {Alijani}}, \bibinfo {author} {\bibfnamefont {M.}~\bibnamefont {Xu}}, \bibinfo {author} {\bibfnamefont {R.~A.}\ \bibnamefont {Norte}}, \ and\ \bibinfo {author} {\bibfnamefont {P.~G.}\ \bibnamefont {Steeneken}},\ }\bibfield  {title} {\enquote {\bibinfo {title} {Resolution limits of resonant sensors},}\ }\href@noop {} {\bibfield  {journal} {\bibinfo  {journal} {Physical Review Applied}\ }\textbf {\bibinfo {volume} {19}},\ \bibinfo {pages} {054074} (\bibinfo {year} {2023})}\BibitemShut {NoStop}%
\bibitem [{\citenamefont {Catalini}\ \emph {et~al.}(2021)\citenamefont {Catalini}, \citenamefont {Rossi}, \citenamefont {Langman},\ and\ \citenamefont {Schliesser}}]{catalini2021modeling}%
  \BibitemOpen
  \bibfield  {author} {\bibinfo {author} {\bibfnamefont {L.}~\bibnamefont {Catalini}}, \bibinfo {author} {\bibfnamefont {M.}~\bibnamefont {Rossi}}, \bibinfo {author} {\bibfnamefont {E.~C.}\ \bibnamefont {Langman}}, \ and\ \bibinfo {author} {\bibfnamefont {A.}~\bibnamefont {Schliesser}},\ }\bibfield  {title} {\enquote {\bibinfo {title} {Modeling and observation of nonlinear damping in dissipation-diluted nanomechanical resonators},}\ }\href@noop {} {\bibfield  {journal} {\bibinfo  {journal} {Physical Review Letters}\ }\textbf {\bibinfo {volume} {126}},\ \bibinfo {pages} {174101} (\bibinfo {year} {2021})}\BibitemShut {NoStop}%
\bibitem [{\citenamefont {Rogalski}(2002)}]{rogalski2002infrared}%
  \BibitemOpen
  \bibfield  {author} {\bibinfo {author} {\bibfnamefont {A.}~\bibnamefont {Rogalski}},\ }\bibfield  {title} {\enquote {\bibinfo {title} {Infrared detectors: an overview},}\ }\href@noop {} {\bibfield  {journal} {\bibinfo  {journal} {Infrared physics \& technology}\ }\textbf {\bibinfo {volume} {43}},\ \bibinfo {pages} {187--210} (\bibinfo {year} {2002})}\BibitemShut {NoStop}%
\bibitem [{\citenamefont {Jones}(1953)}]{jones1953performance}%
  \BibitemOpen
  \bibfield  {author} {\bibinfo {author} {\bibfnamefont {R.~C.}\ \bibnamefont {Jones}},\ }\bibfield  {title} {\enquote {\bibinfo {title} {Performance of detectors for visible and infrared radiation},}\ }\href@noop {} {\bibfield  {journal} {\bibinfo  {journal} {Advances in Electronics and Electron Physics}\ }\textbf {\bibinfo {volume} {5}},\ \bibinfo {pages} {1--96} (\bibinfo {year} {1953})}\BibitemShut {NoStop}%
\bibitem [{\citenamefont {Nudelman}(1962)}]{nudelman1962detectivity}%
  \BibitemOpen
  \bibfield  {author} {\bibinfo {author} {\bibfnamefont {S.}~\bibnamefont {Nudelman}},\ }\bibfield  {title} {\enquote {\bibinfo {title} {The detectivity of infrared photodetectors},}\ }\href@noop {} {\bibfield  {journal} {\bibinfo  {journal} {Applied Optics}\ }\textbf {\bibinfo {volume} {1}},\ \bibinfo {pages} {627--636} (\bibinfo {year} {1962})}\BibitemShut {NoStop}%
\bibitem [{\citenamefont {Zhao}\ \emph {et~al.}(2002)\citenamefont {Zhao}, \citenamefont {Mao}, \citenamefont {Horowitz}, \citenamefont {Majumdar}, \citenamefont {Varesi}, \citenamefont {Norton},\ and\ \citenamefont {Kitching}}]{zhao2002optomechanical}%
  \BibitemOpen
  \bibfield  {author} {\bibinfo {author} {\bibfnamefont {Y.}~\bibnamefont {Zhao}}, \bibinfo {author} {\bibfnamefont {M.}~\bibnamefont {Mao}}, \bibinfo {author} {\bibfnamefont {R.}~\bibnamefont {Horowitz}}, \bibinfo {author} {\bibfnamefont {A.}~\bibnamefont {Majumdar}}, \bibinfo {author} {\bibfnamefont {J.}~\bibnamefont {Varesi}}, \bibinfo {author} {\bibfnamefont {P.}~\bibnamefont {Norton}}, \ and\ \bibinfo {author} {\bibfnamefont {J.}~\bibnamefont {Kitching}},\ }\bibfield  {title} {\enquote {\bibinfo {title} {Optomechanical uncooled infrared imaging system: design, microfabrication, and performance},}\ }\href@noop {} {\bibfield  {journal} {\bibinfo  {journal} {Journal of microelectromechanical systems}\ }\textbf {\bibinfo {volume} {11}},\ \bibinfo {pages} {136--146} (\bibinfo {year} {2002})}\BibitemShut {NoStop}%
\bibitem [{\citenamefont {Senesac}\ \emph {et~al.}(2003)\citenamefont {Senesac}, \citenamefont {Corbeil}, \citenamefont {Rajic}, \citenamefont {Lavrik},\ and\ \citenamefont {Datskos}}]{senesac2003ir}%
  \BibitemOpen
  \bibfield  {author} {\bibinfo {author} {\bibfnamefont {L.}~\bibnamefont {Senesac}}, \bibinfo {author} {\bibfnamefont {J.}~\bibnamefont {Corbeil}}, \bibinfo {author} {\bibfnamefont {S.}~\bibnamefont {Rajic}}, \bibinfo {author} {\bibfnamefont {N.}~\bibnamefont {Lavrik}}, \ and\ \bibinfo {author} {\bibfnamefont {P.}~\bibnamefont {Datskos}},\ }\bibfield  {title} {\enquote {\bibinfo {title} {Ir imaging using uncooled microcantilever detectors},}\ }\href@noop {} {\bibfield  {journal} {\bibinfo  {journal} {Ultramicroscopy}\ }\textbf {\bibinfo {volume} {97}},\ \bibinfo {pages} {451--458} (\bibinfo {year} {2003})}\BibitemShut {NoStop}%
\bibitem [{\citenamefont {Gokhale}\ \emph {et~al.}(2015)\citenamefont {Gokhale}, \citenamefont {Figueroa}, \citenamefont {Tsai},\ and\ \citenamefont {Rais-Zadeh}}]{gokhale2015low}%
  \BibitemOpen
  \bibfield  {author} {\bibinfo {author} {\bibfnamefont {V.~J.}\ \bibnamefont {Gokhale}}, \bibinfo {author} {\bibfnamefont {C.}~\bibnamefont {Figueroa}}, \bibinfo {author} {\bibfnamefont {J.~M.~L.}\ \bibnamefont {Tsai}}, \ and\ \bibinfo {author} {\bibfnamefont {M.}~\bibnamefont {Rais-Zadeh}},\ }\bibfield  {title} {\enquote {\bibinfo {title} {Low-noise aln-on-si resonant infrared detectors using a commercial foundry mems fabrication process},}\ }in\ \href@noop {} {\emph {\bibinfo {booktitle} {2015 28th IEEE International Conference on Micro Electro Mechanical Systems (MEMS)}}}\ (\bibinfo {organization} {IEEE},\ \bibinfo {year} {2015})\ pp.\ \bibinfo {pages} {73--76}\BibitemShut {NoStop}%
\bibitem [{\citenamefont {Kanellopulos}\ \emph {et~al.}(2024)\citenamefont {Kanellopulos}, \citenamefont {West}, \citenamefont {Emminger}, \citenamefont {Martini}, \citenamefont {Sauer}, \citenamefont {Foelske},\ and\ \citenamefont {Schmid}}]{kanellopulos2024stress}%
  \BibitemOpen
  \bibfield  {author} {\bibinfo {author} {\bibfnamefont {K.}~\bibnamefont {Kanellopulos}}, \bibinfo {author} {\bibfnamefont {R.~G.}\ \bibnamefont {West}}, \bibinfo {author} {\bibfnamefont {S.}~\bibnamefont {Emminger}}, \bibinfo {author} {\bibfnamefont {P.}~\bibnamefont {Martini}}, \bibinfo {author} {\bibfnamefont {M.}~\bibnamefont {Sauer}}, \bibinfo {author} {\bibfnamefont {A.}~\bibnamefont {Foelske}}, \ and\ \bibinfo {author} {\bibfnamefont {S.}~\bibnamefont {Schmid}},\ }\bibfield  {title} {\enquote {\bibinfo {title} {Stress-dependent optical extinction in low-pressure chemical vapor deposition silicon nitride measured by nanomechanical photothermal sensing},}\ }\href@noop {} {\bibfield  {journal} {\bibinfo  {journal} {Nano Letters}\ }\textbf {\bibinfo {volume} {24}},\ \bibinfo {pages} {11262--11268} (\bibinfo {year} {2024})}\BibitemShut {NoStop}%
\bibitem [{\citenamefont {Toivola}\ \emph {et~al.}(2003)\citenamefont {Toivola}, \citenamefont {Thurn}, \citenamefont {Cook}, \citenamefont {Cibuzar},\ and\ \citenamefont {Roberts}}]{toivola2003influence}%
  \BibitemOpen
  \bibfield  {author} {\bibinfo {author} {\bibfnamefont {Y.}~\bibnamefont {Toivola}}, \bibinfo {author} {\bibfnamefont {J.}~\bibnamefont {Thurn}}, \bibinfo {author} {\bibfnamefont {R.~F.}\ \bibnamefont {Cook}}, \bibinfo {author} {\bibfnamefont {G.}~\bibnamefont {Cibuzar}}, \ and\ \bibinfo {author} {\bibfnamefont {K.}~\bibnamefont {Roberts}},\ }\bibfield  {title} {\enquote {\bibinfo {title} {Influence of deposition conditions on mechanical properties of low-pressure chemical vapor deposited low-stress silicon nitride films},}\ }\href@noop {} {\bibfield  {journal} {\bibinfo  {journal} {Journal of applied physics}\ }\textbf {\bibinfo {volume} {94}},\ \bibinfo {pages} {6915--6922} (\bibinfo {year} {2003})}\BibitemShut {NoStop}%
\bibitem [{\citenamefont {Ftouni}\ \emph {et~al.}(2015)\citenamefont {Ftouni}, \citenamefont {Blanc}, \citenamefont {Tainoff}, \citenamefont {Fefferman}, \citenamefont {Defoort}, \citenamefont {Lulla}, \citenamefont {Richard}, \citenamefont {Collin},\ and\ \citenamefont {Bourgeois}}]{ftouni2015thermal}%
  \BibitemOpen
  \bibfield  {author} {\bibinfo {author} {\bibfnamefont {H.}~\bibnamefont {Ftouni}}, \bibinfo {author} {\bibfnamefont {C.}~\bibnamefont {Blanc}}, \bibinfo {author} {\bibfnamefont {D.}~\bibnamefont {Tainoff}}, \bibinfo {author} {\bibfnamefont {A.~D.}\ \bibnamefont {Fefferman}}, \bibinfo {author} {\bibfnamefont {M.}~\bibnamefont {Defoort}}, \bibinfo {author} {\bibfnamefont {K.~J.}\ \bibnamefont {Lulla}}, \bibinfo {author} {\bibfnamefont {J.}~\bibnamefont {Richard}}, \bibinfo {author} {\bibfnamefont {E.}~\bibnamefont {Collin}}, \ and\ \bibinfo {author} {\bibfnamefont {O.}~\bibnamefont {Bourgeois}},\ }\bibfield  {title} {\enquote {\bibinfo {title} {Thermal conductivity of silicon nitride membranes is not sensitive to stress},}\ }\href@noop {} {\bibfield  {journal} {\bibinfo  {journal} {Physical Review B}\ }\textbf {\bibinfo {volume} {92}},\ \bibinfo {pages} {125439} (\bibinfo {year} {2015})}\BibitemShut {NoStop}%
\bibitem [{\citenamefont {Hilsum}(1954)}]{hilsum1954absorption}%
  \BibitemOpen
  \bibfield  {author} {\bibinfo {author} {\bibfnamefont {C.}~\bibnamefont {Hilsum}},\ }\bibfield  {title} {\enquote {\bibinfo {title} {Infrared absorption of thin metal films},}\ }\href {\doibase 10.1364/JOSA.44.000188} {\bibfield  {journal} {\bibinfo  {journal} {J. Opt. Soc. Am.}\ }\textbf {\bibinfo {volume} {44}},\ \bibinfo {pages} {188--191} (\bibinfo {year} {1954})}\BibitemShut {NoStop}%
\bibitem [{\citenamefont {Luhmann}\ \emph {et~al.}(2020)\citenamefont {Luhmann}, \citenamefont {H{\o}j}, \citenamefont {Piller}, \citenamefont {K{\"a}hler}, \citenamefont {Chien}, \citenamefont {West}, \citenamefont {Andersen},\ and\ \citenamefont {Schmid}}]{luhmann2020ultrathin}%
  \BibitemOpen
  \bibfield  {author} {\bibinfo {author} {\bibfnamefont {N.}~\bibnamefont {Luhmann}}, \bibinfo {author} {\bibfnamefont {D.}~\bibnamefont {H{\o}j}}, \bibinfo {author} {\bibfnamefont {M.}~\bibnamefont {Piller}}, \bibinfo {author} {\bibfnamefont {H.}~\bibnamefont {K{\"a}hler}}, \bibinfo {author} {\bibfnamefont {M.-H.}\ \bibnamefont {Chien}}, \bibinfo {author} {\bibfnamefont {R.~G.}\ \bibnamefont {West}}, \bibinfo {author} {\bibfnamefont {U.~L.}\ \bibnamefont {Andersen}}, \ and\ \bibinfo {author} {\bibfnamefont {S.}~\bibnamefont {Schmid}},\ }\bibfield  {title} {\enquote {\bibinfo {title} {Ultrathin 2 nm gold as impedance-matched absorber for infrared light},}\ }\href@noop {} {\bibfield  {journal} {\bibinfo  {journal} {Nature communications}\ }\textbf {\bibinfo {volume} {11}},\ \bibinfo {pages} {2161} (\bibinfo {year} {2020})}\BibitemShut {NoStop}%
\bibitem [{\citenamefont {Edalatpour}\ and\ \citenamefont {Francoeur}(2013)}]{edalatpour2013size}%
  \BibitemOpen
  \bibfield  {author} {\bibinfo {author} {\bibfnamefont {S.}~\bibnamefont {Edalatpour}}\ and\ \bibinfo {author} {\bibfnamefont {M.}~\bibnamefont {Francoeur}},\ }\bibfield  {title} {\enquote {\bibinfo {title} {Size effect on the emissivity of thin films},}\ }\href@noop {} {\bibfield  {journal} {\bibinfo  {journal} {Journal of Quantitative Spectroscopy and Radiative Transfer}\ }\textbf {\bibinfo {volume} {118}},\ \bibinfo {pages} {75--85} (\bibinfo {year} {2013})}\BibitemShut {NoStop}%
\bibitem [{\citenamefont {Bergman}(2011)}]{bergman2011fundamentals}%
  \BibitemOpen
  \bibfield  {author} {\bibinfo {author} {\bibfnamefont {T.~L.}\ \bibnamefont {Bergman}},\ }\href@noop {} {\emph {\bibinfo {title} {Fundamentals of heat and mass transfer}}}\ (\bibinfo  {publisher} {John Wiley \& Sons},\ \bibinfo {year} {2011})\BibitemShut {NoStop}%
\bibitem [{\citenamefont {Cleland}\ and\ \citenamefont {Roukes}(2002)}]{cleland2002noise}%
  \BibitemOpen
  \bibfield  {author} {\bibinfo {author} {\bibfnamefont {A.}~\bibnamefont {Cleland}}\ and\ \bibinfo {author} {\bibfnamefont {M.}~\bibnamefont {Roukes}},\ }\bibfield  {title} {\enquote {\bibinfo {title} {Noise processes in nanomechanical resonators},}\ }\href@noop {} {\bibfield  {journal} {\bibinfo  {journal} {Journal of applied physics}\ }\textbf {\bibinfo {volume} {92}},\ \bibinfo {pages} {2758--2769} (\bibinfo {year} {2002})}\BibitemShut {NoStop}%
\bibitem [{\citenamefont {Djuri{\'c}}, \citenamefont {Jak{\v{s}}i{\'c}},\ and\ \citenamefont {Randjelovi{\'c}}(2002)}]{djuric2002adsorption}%
  \BibitemOpen
  \bibfield  {author} {\bibinfo {author} {\bibfnamefont {Z.}~\bibnamefont {Djuri{\'c}}}, \bibinfo {author} {\bibfnamefont {O.}~\bibnamefont {Jak{\v{s}}i{\'c}}}, \ and\ \bibinfo {author} {\bibfnamefont {D.}~\bibnamefont {Randjelovi{\'c}}},\ }\bibfield  {title} {\enquote {\bibinfo {title} {Adsorption--desorption noise in micromechanical resonant structures},}\ }\href@noop {} {\bibfield  {journal} {\bibinfo  {journal} {Sensors and Actuators A: Physical}\ }\textbf {\bibinfo {volume} {96}},\ \bibinfo {pages} {244--251} (\bibinfo {year} {2002})}\BibitemShut {NoStop}%
\bibitem [{\citenamefont {Fong}, \citenamefont {Pernice},\ and\ \citenamefont {Tang}(2012)}]{fong2012frequency}%
  \BibitemOpen
  \bibfield  {author} {\bibinfo {author} {\bibfnamefont {K.~Y.}\ \bibnamefont {Fong}}, \bibinfo {author} {\bibfnamefont {W.~H.}\ \bibnamefont {Pernice}}, \ and\ \bibinfo {author} {\bibfnamefont {H.~X.}\ \bibnamefont {Tang}},\ }\bibfield  {title} {\enquote {\bibinfo {title} {Frequency and phase noise of ultrahigh q silicon nitride nanomechanical resonators},}\ }\href@noop {} {\bibfield  {journal} {\bibinfo  {journal} {Physical Review B—Condensed Matter and Materials Physics}\ }\textbf {\bibinfo {volume} {85}},\ \bibinfo {pages} {161410} (\bibinfo {year} {2012})}\BibitemShut {NoStop}%
\bibitem [{\citenamefont {Zhang}\ and\ \citenamefont {St-Gelais}(2023)}]{zhang2023demonstration}%
  \BibitemOpen
  \bibfield  {author} {\bibinfo {author} {\bibfnamefont {C.}~\bibnamefont {Zhang}}\ and\ \bibinfo {author} {\bibfnamefont {R.}~\bibnamefont {St-Gelais}},\ }\bibfield  {title} {\enquote {\bibinfo {title} {Demonstration of frequency stability limited by thermal fluctuation noise in silicon nitride nanomechanical resonators},}\ }\href@noop {} {\bibfield  {journal} {\bibinfo  {journal} {Applied Physics Letters}\ }\textbf {\bibinfo {volume} {122}} (\bibinfo {year} {2023})}\BibitemShut {NoStop}%
\bibitem [{\citenamefont {Moura}\ \emph {et~al.}(2018)\citenamefont {Moura}, \citenamefont {Norte}, \citenamefont {Guo}, \citenamefont {Sch{\"a}fermeier},\ and\ \citenamefont {Gr{\"o}blacher}}]{moura2018centimeter}%
  \BibitemOpen
  \bibfield  {author} {\bibinfo {author} {\bibfnamefont {J.~P.}\ \bibnamefont {Moura}}, \bibinfo {author} {\bibfnamefont {R.~A.}\ \bibnamefont {Norte}}, \bibinfo {author} {\bibfnamefont {J.}~\bibnamefont {Guo}}, \bibinfo {author} {\bibfnamefont {C.}~\bibnamefont {Sch{\"a}fermeier}}, \ and\ \bibinfo {author} {\bibfnamefont {S.}~\bibnamefont {Gr{\"o}blacher}},\ }\bibfield  {title} {\enquote {\bibinfo {title} {Centimeter-scale suspended photonic crystal mirrors},}\ }\href@noop {} {\bibfield  {journal} {\bibinfo  {journal} {Optics express}\ }\textbf {\bibinfo {volume} {26}},\ \bibinfo {pages} {1895--1909} (\bibinfo {year} {2018})}\BibitemShut {NoStop}%
\bibitem [{\citenamefont {Norder}\ \emph {et~al.}(2025)\citenamefont {Norder}, \citenamefont {Yin}, \citenamefont {de~Jong}, \citenamefont {Stallone}, \citenamefont {Aydogmus}, \citenamefont {Sberna}, \citenamefont {Bessa},\ and\ \citenamefont {Norte}}]{norder2025pentagonal}%
  \BibitemOpen
  \bibfield  {author} {\bibinfo {author} {\bibfnamefont {L.}~\bibnamefont {Norder}}, \bibinfo {author} {\bibfnamefont {S.}~\bibnamefont {Yin}}, \bibinfo {author} {\bibfnamefont {M.~H.}\ \bibnamefont {de~Jong}}, \bibinfo {author} {\bibfnamefont {F.}~\bibnamefont {Stallone}}, \bibinfo {author} {\bibfnamefont {H.}~\bibnamefont {Aydogmus}}, \bibinfo {author} {\bibfnamefont {P.~M.}\ \bibnamefont {Sberna}}, \bibinfo {author} {\bibfnamefont {M.~A.}\ \bibnamefont {Bessa}}, \ and\ \bibinfo {author} {\bibfnamefont {R.~A.}\ \bibnamefont {Norte}},\ }\bibfield  {title} {\enquote {\bibinfo {title} {Pentagonal photonic crystal mirrors: scalable lightsails with enhanced acceleration via neural topology optimization},}\ }\href@noop {} {\bibfield  {journal} {\bibinfo  {journal} {Nature Communications}\ }\textbf {\bibinfo {volume} {16}},\ \bibinfo {pages} {2753} (\bibinfo {year} {2025})}\BibitemShut {NoStop}%
\bibitem [{\citenamefont {Dutt}\ \emph {et~al.}(2023)\citenamefont {Dutt}, \citenamefont {Karawdeniya}, \citenamefont {Bandara}, \citenamefont {Afrin},\ and\ \citenamefont {Kluth}}]{dutt2023ultrathin}%
  \BibitemOpen
  \bibfield  {author} {\bibinfo {author} {\bibfnamefont {S.}~\bibnamefont {Dutt}}, \bibinfo {author} {\bibfnamefont {B.~I.}\ \bibnamefont {Karawdeniya}}, \bibinfo {author} {\bibfnamefont {Y.~N.~D.}\ \bibnamefont {Bandara}}, \bibinfo {author} {\bibfnamefont {N.}~\bibnamefont {Afrin}}, \ and\ \bibinfo {author} {\bibfnamefont {P.}~\bibnamefont {Kluth}},\ }\bibfield  {title} {\enquote {\bibinfo {title} {Ultrathin, high-lifetime silicon nitride membranes for nanopore sensing},}\ }\href@noop {} {\bibfield  {journal} {\bibinfo  {journal} {Analytical Chemistry}\ }\textbf {\bibinfo {volume} {95}},\ \bibinfo {pages} {5754--5763} (\bibinfo {year} {2023})}\BibitemShut {NoStop}%
\bibitem [{\citenamefont {Tong}\ \emph {et~al.}(2004)\citenamefont {Tong}, \citenamefont {Jansen}, \citenamefont {Gadgil}, \citenamefont {Bostan}, \citenamefont {Berenschot}, \citenamefont {van Rijn},\ and\ \citenamefont {Elwenspoek}}]{tong2004silicon}%
  \BibitemOpen
  \bibfield  {author} {\bibinfo {author} {\bibfnamefont {H.~D.}\ \bibnamefont {Tong}}, \bibinfo {author} {\bibfnamefont {H.~V.}\ \bibnamefont {Jansen}}, \bibinfo {author} {\bibfnamefont {V.~J.}\ \bibnamefont {Gadgil}}, \bibinfo {author} {\bibfnamefont {C.~G.}\ \bibnamefont {Bostan}}, \bibinfo {author} {\bibfnamefont {E.}~\bibnamefont {Berenschot}}, \bibinfo {author} {\bibfnamefont {C.~J.}\ \bibnamefont {van Rijn}}, \ and\ \bibinfo {author} {\bibfnamefont {M.}~\bibnamefont {Elwenspoek}},\ }\bibfield  {title} {\enquote {\bibinfo {title} {Silicon nitride nanosieve membrane},}\ }\href@noop {} {\bibfield  {journal} {\bibinfo  {journal} {Nano letters}\ }\textbf {\bibinfo {volume} {4}},\ \bibinfo {pages} {283--287} (\bibinfo {year} {2004})}\BibitemShut {NoStop}%
\end{thebibliography}%

\end{document}



\title{Supplementary Information: Design optimisation of silicon nitride nanomechanical resonators for thermal infrared detectors: a guide through key figures of merit}



\author{Paolo Martini}
\affiliation{Institute of Sensor and Actuator Systems, TU Wien, Gusshaustrasse 27-29, Vienna, 1040, Austria}
\author{Kostas Kanellopulos}
\affiliation{Institute of Sensor and Actuator Systems, TU Wien, Gusshaustrasse 27-29, Vienna, 1040, Austria}
\author{Silvan Schmid}
 \email{silvan.schmid@tuwien.ac.at}
\affiliation{Institute of Sensor and Actuator Systems, TU Wien, Gusshaustrasse 27-29, Vienna, 1040, Austria}


\date{\today}

\pacs{}

\maketitle 

\section{Duffing parameter of trampolines}
\label{S_sec:duffing}
\begin{figure}
    \centering
    \includegraphics[width=1\linewidth]{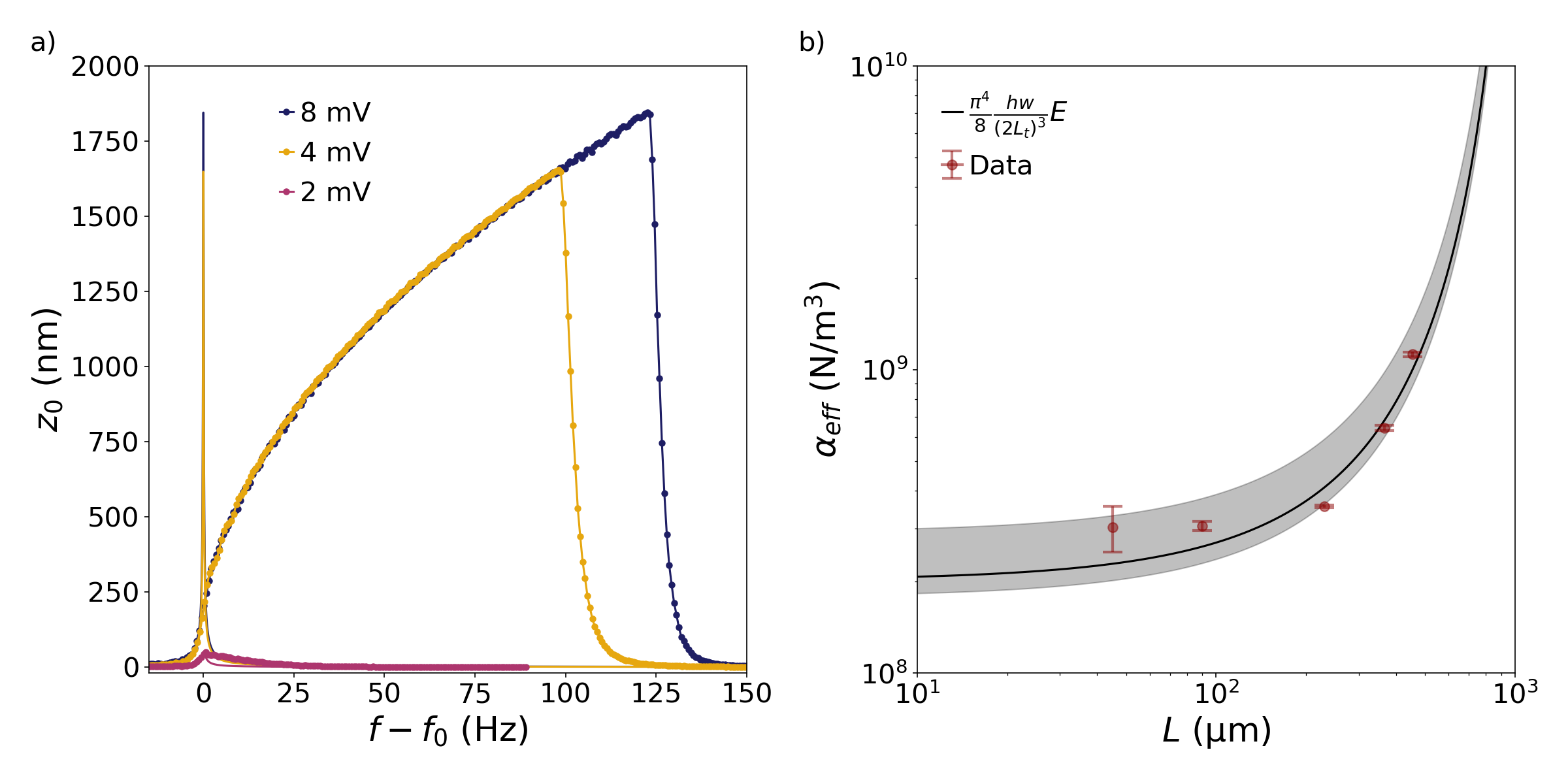}
    \caption{Duffing parameter of trampolines. a) Non-linear responses (shark-fin shaped curves) and linear responses (Lorentzian peak curves) obtained after the fitting with Eq. \ref{S_eq:linear_response}. b) Comparison between the measurements (red dots) and theory (black line). The grey band represents the uncertainty deriving from the material parameters.}
    \label{S_fig:duffing}
\end{figure}

We verify the formulation presented in Equation (20) in the main text using the method described in \cite{li2025finite}. The amplitude response of trampoline resonators was recorded at different driving amplitudes while sweeping the driving frequency. From the resulting nonlinear response (shark-fin shaped curves in Fig. \ref{S_fig:duffing}a), we extracted the peak amplitude $z_{0,nl}$ and the corresponding frequency $f_{max,nl}$. To correct for frequency drifts during the sweeps, the off-resonance regions of the data were fitted with the linear amplitude response \cite{li2025finite}:

\begin{equation}
    z_0(f)=\frac{z_{0,nl}/Q}{\sqrt{[1-(f/f_0)^2]^2+f^2/(f_0Q)^2}}
    \label{S_eq:linear_response}
\end{equation}
yielding the linear resonance frequency, $f_0$. We then replotted the response curves along the shifted frequency axis $f-f_0$ (Figure \ref{S_fig:duffing}a), and fitted the backbone of the points ($z_{0,nl}$,$f_{max,nl}$) with \cite{li2025finite}:

\begin{equation}
    f^2_{max,nl}=f^2_0+\frac{3}{16\pi^2}\frac{\alpha_{Duff,t}}{m_eff}z^2_{0,nl}
\end{equation}
to extract the non-linear duffing parameters, $\alpha_{Duff,t}$.\newline
The results are presented in figure \ref{S_fig:duffing}b, for trampolines of different sizes. Varying the central area side length $L$ while keeping the frame fixed changes the tether length, which in turn affects the Duffing parameter. The experimental data (red dots) are in good agreement with the theoretical prediction (black line). The grey band represents the uncertainty arising from the material parameters. 

\section{Tethers width, $w$}
\label{sec:w}

\begin{figure}
    \centering
    \includegraphics[width=1\linewidth]{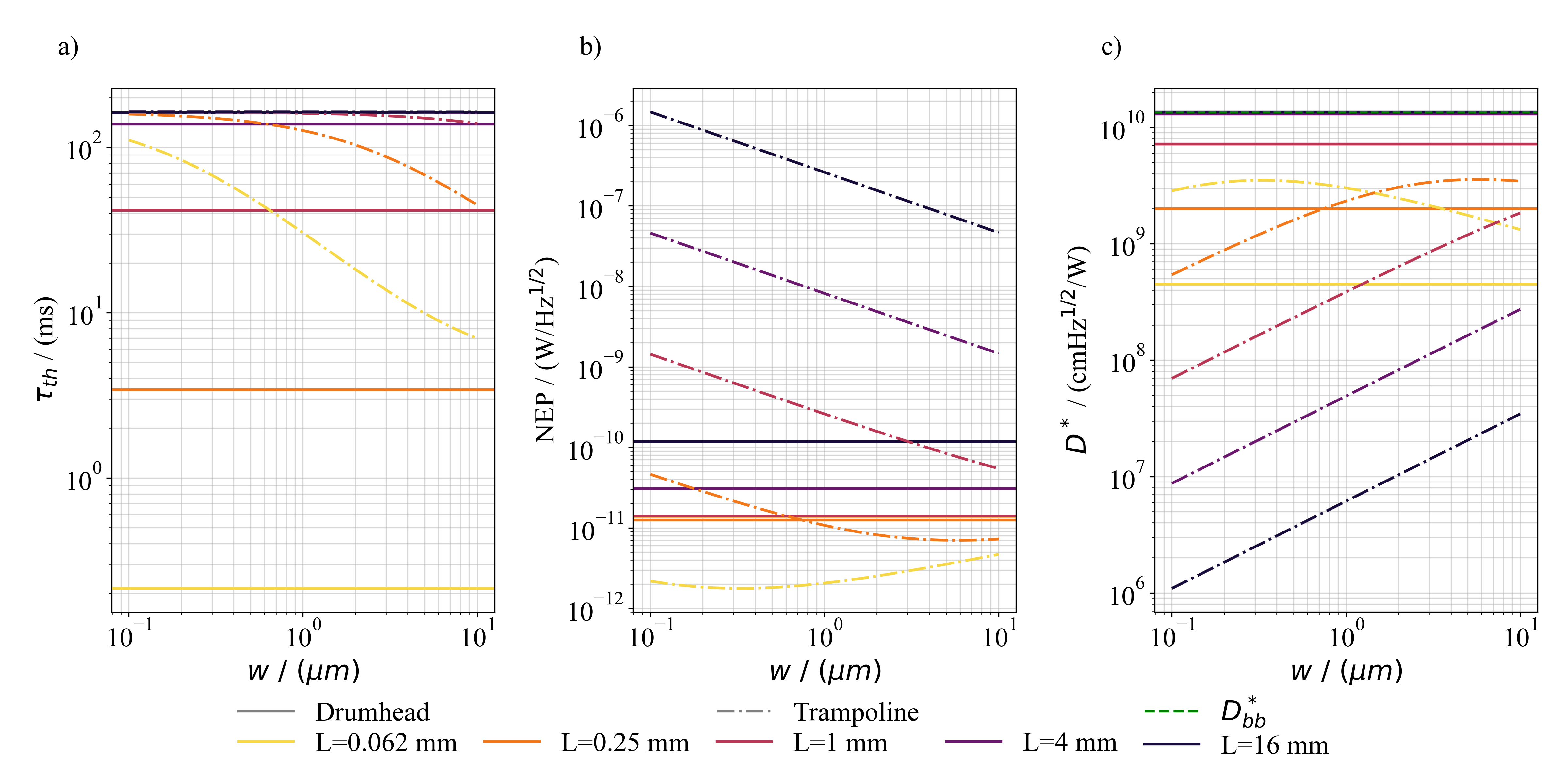}
    \caption{Influence of the tether width $w$ on the three figures of merit: a) thermal time constant $\tau_{th}$; b) noise equivalent power (NEP); c) specific detectivity $D^*$ at $\lambda_{SiN} = \SI{11.86}{\micro\m}$ for plain SiN resonators without a dedicated IR absorber. The dashed black line in c) marks the room-temperature sensitivity limit. For all panels, the stress is fixed at $\sigma = \SI{200}{\mega\pascal}$, the thickness $h = \SI{50}{\nano\m}$, and the tether width at $w = \SI{5}{\micro\m}$.}
    \label{fig:3in1_w_SiN}
\end{figure}
As with the tether length, the tether width influences both the thermal conductance $G$ and the heat capacitance $C$. Wider tethers facilitate heat dissipation via conduction and simultaneously increase the overall radiative surface area. Higher values of $w$ also correspond to a larger volume available for thermal storage, thereby increasing $C$. The resulting behaviour of NEP and $D^*$ as a function of $w$ closely mirrors that observed for $L_t$. 

Once again, this variable concerns only trampolines. Therefore, results form drumheads are presented for reference as constant, horizontal lines in all plots of Figure \ref{fig:3in1_w_SiN}. The thickness is fixed at $h=\SI{50}{\nano\m}$, the tethers length at $L_t=\SI{50}{\micro\m}$, and the stress at $\sigma = \SI{200}{\mega\pascal}$. The effect of varying $w$ in the range \SIrange{0.1}{10}{\micro\m} on $\tau_{th}$ is illustrated in Figure \ref{fig:3in1_w_SiN}a. In general, structures with wider tethers dissipate heat more efficiently, leading to a lower thermal time constant. As observed for $L_t$, both trampolines and drumheads converge to the same asymptotic value of $\tau_{th}$ as they transition into the radiative heat transfer regime. While for the tethers' length this happened for longer tethers, for $w$ this occurs as the tethers become thinner. 
Overall, the presence of tethers and the increased thermal isolation they introduce make trampolines generally slower than drumheads of equivalent side length $L$.

Next is the noise equivalent power, shown in Figure \ref{fig:3in1_w_SiN}b. Once again, the best-performing structure is the trampoline with $L=\SI{0.062}{\micro\m}$, which exhibits the lowest NEP at $w\simeq \SI{300}{\nano\m}$. 
Trampolines with larger detection areas are generally outperformed by drumheads of the same size. This performance gap is primarily due to differences in frequency stability, $S_{y_{\theta}}$.

Finally, when examining the specific detectivity $D^*$ in Figure \ref{fig:3in1_w_SiN}c, we observe that generally bigger drumheads exhibit higher detectivity values than trampolines. Interestingly, trampolines with different detection area sizes show very similar $D^*$ performance at their respective maxima. However, the position of these maxima varies with the detection area side length: the trampoline with $L=\SI{0.062}{\milli\m}$ peaks at $w=\SI{400}{\nano\m}$, the one with $L=\SI{0.25}{\milli\m}$ peaks at $w=\SI{5}{\micro\m}$.

\begin{figure}
    \centering
    \includegraphics[width=1\linewidth]{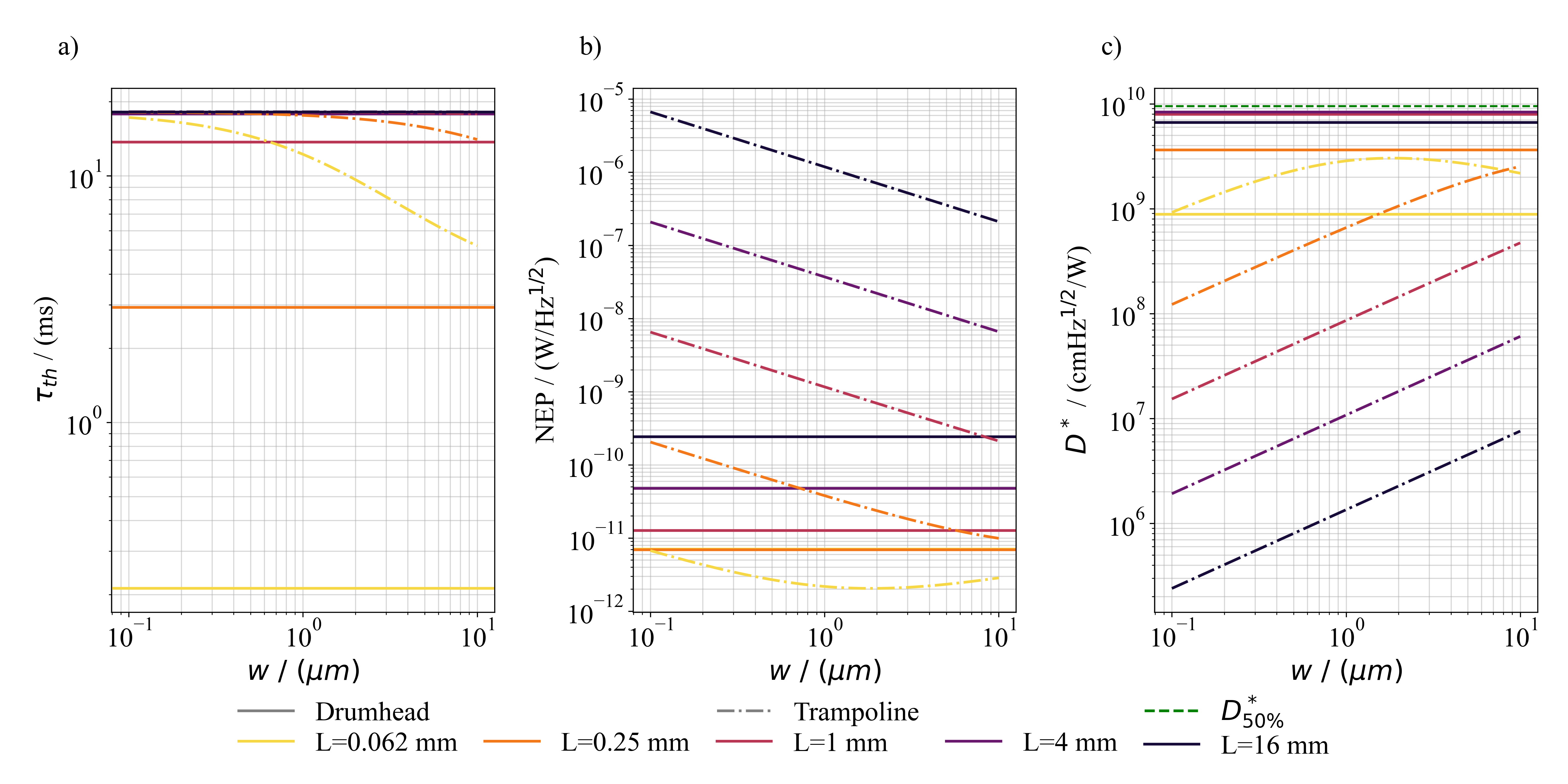}
    \caption{Influence of the tether width $w$ on the three figures of merit: a) thermal time constant $\tau_{th}$; b) noise equivalent power (NEP); c) specific detectivity $D^*$ for a SiN resonator with an IR absorber. The dashed black line in c) marks the room-temperature sensitivity limit for $\alpha_{abs}=50\%$. For all panels, the stress is fixed at $\sigma = \SI{200}{\mega\pascal}$, the thickness $h = \SI{50}{\nano\m}$, and the tether width at $w = \SI{5}{\micro\m}$.}
    \label{fig:3in1_w_abs50}
\end{figure}
When comparing the results for bare SiN detectors (Figure \ref{fig:3in1_w_SiN}) with those for detectors incorporating a metal IR absorber layer (Figure \ref{fig:3in1_w_abs50}), a clear reduction in the thermal time constant is observed (Figure \ref{fig:3in1_w_abs50}a). In addition, the optimal values of both $NEP$ and $D^*$ (Figure \ref{fig:3in1_w_abs50}b and \ref{fig:3in1_w_abs50}c, respectively) shift toward larger $w$ for all trampoline dimensions. Both effects can be attributed to the enhanced thermal properties introduced by the metal layer.

\section{Absorptance and emissivity of SiN at varying thicknesses $h$}
To calculate the absorptivity and emissivity of different thicknesses of SiN, we first used the Python interface pySCATMECH \cite{germer2020pyscatmech}. We plugged in the complex refractive index:

\begin{equation}
    \bar{n}=n - i\kappa
\end{equation}
with $n$ and $\kappa$ the refractive index and extinction coefficient, respectively, as obtained by \textit{Cataldo et al.} \cite{cataldo2012infrared}. We readily obtained the reflectance $R(h,\lambda)$ and transmittance $T(h,\lambda)$ for our specific case of incident angle $\theta = 0^\circ$. 
From that we obtain the absorptance spectra as:

\begin{equation}
    \alpha_{SiN}(h, \lambda)=1-T(h, \lambda)-R(h, \lambda),
\end{equation}
For each thickness $h$ we proceeded to locate the absorption peak wavelength $\lambda_{SiN}$ and recorded the corresponding absorptance value, $\alpha_{SiN}(\lambda_{SiN})$.

To determine the emissivity, we first computed the total emissive power of a blackbody $E_b$ as \cite{bergman2017fundamentals}:

\begin{equation}
    E_b = \int_{0}^{\infty} \frac{2 \pi h c_0^2}{\lambda^5 \left[\exp\left(\frac{h c_0}{k \lambda T}\right) - 1\right]} \, d\lambda,
    \label{eq:planck_radiation}
\end{equation}
where $h$ is the Planck constant, $k$ is the Boltzmann constant, and $c_0$ the speed of light. The corresponding spectral emissive power $E_{\lambda,b}(\lambda,T)$ was then obtained as following \cite{bergman2017fundamentals}:

\begin{equation}
    E_{\lambda, b}(\lambda, T) = \frac{2 \pi h c_0^2}{\lambda^5 \left[ \exp\left(\frac{h c_0/k}{\lambda T}\right) - 1 \right]}.
\label{eq:planck_law}
\end{equation}

By identifying the absorptance $\alpha$ with the spectral emissivity $\epsilon_{\lambda}(\lambda,T)$ the effective emissivity can be expressed as a weighted average of $\epsilon_{\lambda}$ over the blackbody spectrum:

\begin{equation}
    \varepsilon(T) = \frac{\int_0^{\infty} \varepsilon_{\lambda}(\lambda, T) E_{\lambda, b}(\lambda, T) d\lambda}{E_b(T)}.
    \label{eq:emissivity}
\end{equation}

In our analysis, we set $T=\SI{300}{\kelvin}$ and performed the integration over the wavelength interval $\SIrange[]{1}{667}{\micro\m}$ which yielded an effective emissivity of $\epsilon_{SiN} = 0.06$.

\section{Noise plots}
\begin{figure}
    \centering
    \includegraphics[width=1\linewidth, height=0.95\textheight]{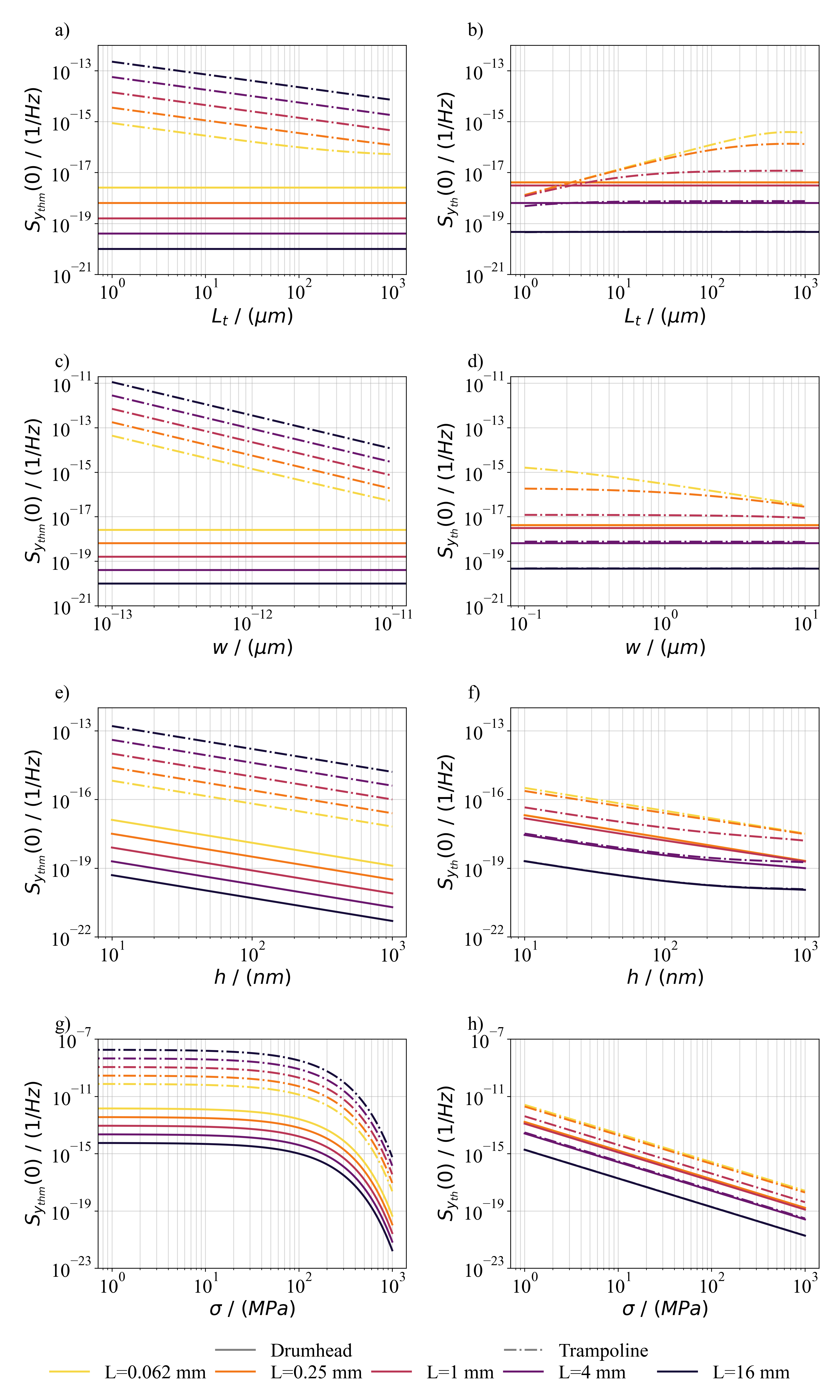}
    \caption{Results of $S_{y_{thm}}(0)$ and $S_{y_{th}}(0)$ for varying a-b) tethers length $L_t$, c-d) tethers width $w$, e-f) thickness $h$, and g-h) tensile stress $\sigma$. For the calculations, we assumed $z_0=z_{0_c}$.}
    \label{S_fig:8in1_noise_SiN}
\end{figure}

In Figure \ref{S_fig:8in1_noise_SiN} we report the influence of noise on SiN drumhead and trampoline resonators as a function of the different geometrical variables and the stress. For each variable, we plot the contribution of the thermomechanical noise, $S_{y_{thm}}(0)$ (Figure \ref{S_fig:8in1_noise_SiN}a, \ref{S_fig:8in1_noise_SiN}c, \ref{S_fig:8in1_noise_SiN}e, and \ref{S_fig:8in1_noise_SiN}g), together with that of the thermal fluctuation noise, $S_{y_{th}}(0)$ (Figure \ref{S_fig:8in1_noise_SiN}b, \ref{S_fig:8in1_noise_SiN}d, \ref{S_fig:8in1_noise_SiN}f, and \ref{S_fig:8in1_noise_SiN}h). Here, we assume that the detection noise is neglegible with respect to the thermomechanical noise, such that $S_{y_\theta}(\omega)\simeq S_{y_{thm}}(\omega)$. Moreover, in the calculations we set $\omega=0$, since the spectral density at zero frequency determines the noise level in the steady state. Evaluating $S_y(0)$ provides a measure of the sensitivity that is independent of the tracking scheme bandwidth and therefore represents a fundamental sensitivity limit of the detector.

Across all variables, $S_{y_{thm}}(0)$ is several orders of magnitude lower for drumheads than for trampolines. This largely explains why, despite the latter generally exhibiting higher power responsivity, the former achieves better $NEP$ and $D^*$. The contribution of $S_{y_{th}}(0)$ further reinforces this trend, being consistently higher for trampolines across almost all geometries and parameter values. For a direct comparison between the two noise sources, the y-axis range was kept identical for each analysed variable.

\section{Bare SiN - supplementary results ($R_P$)}
\begin{figure}
    \centering
    \includegraphics[width=1\linewidth]{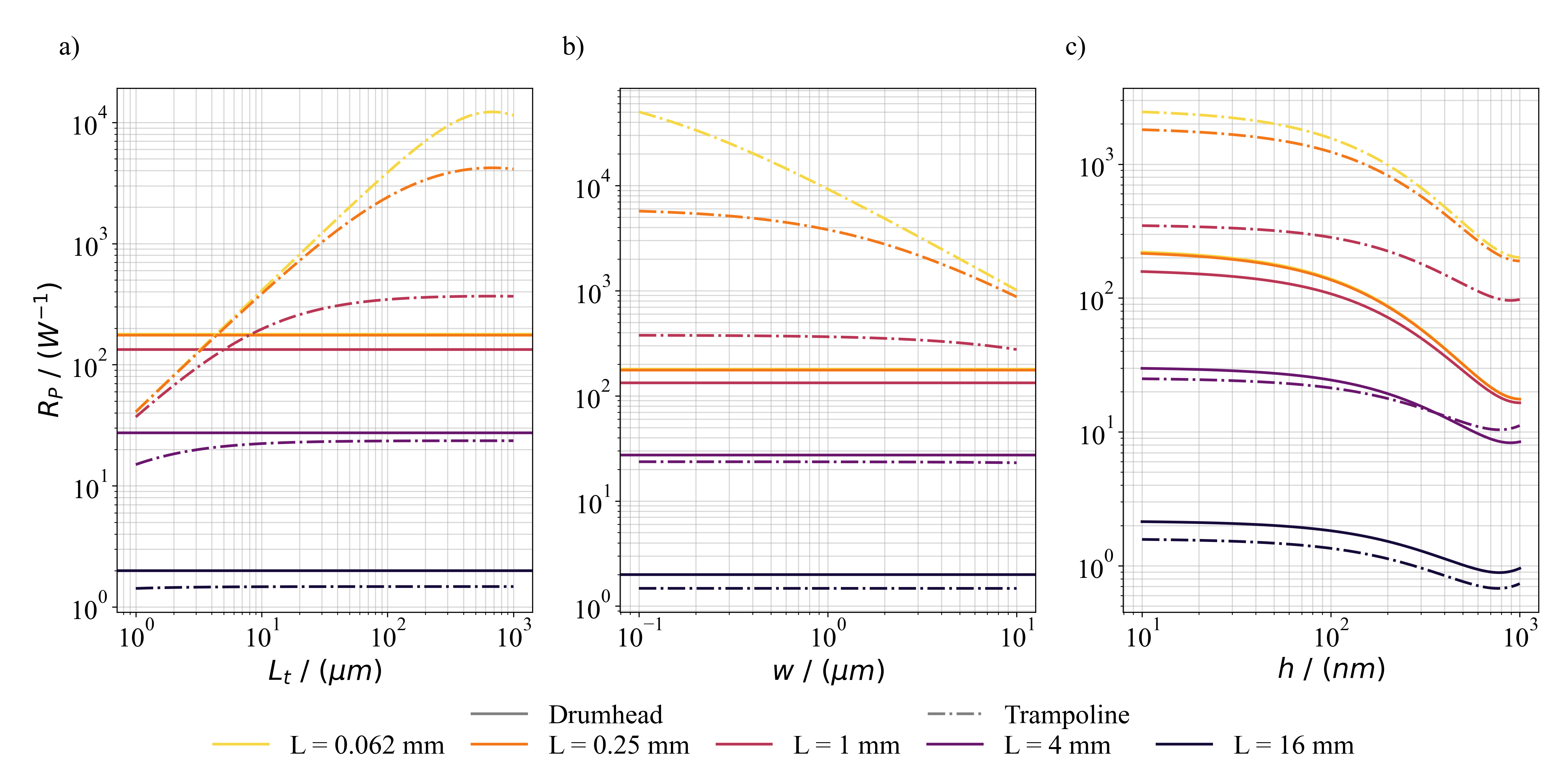}
    \caption{Results for $R_P$ for varying a) tethers length $L_t$, b) tethers width $w$, c) thickness $h$.}
    \label{S_fig:8in1_SiN}
\end{figure}

In figure \ref{S_fig:8in1_SiN}, we present the results for the power responsivity for drumhead and trampoline resonators made of bare SiN. Figure \ref{S_fig:8in1_SiN}a shows the results for varying tether length, where membranes are represented by constant, horizontal lines. The general trend for both drumheads and trampolines is that smaller structures have higher $R_P$. Trampolines have equal or better power responsivity than membranes of equivalent detection area. For trampolines, longer tethers mean increased thermal isolation of the central area, hence higher responsivity. This holds until the structure is fully coupled with the environment through radiation, when $R_P$ then becomes independent of $L_t$. For the tether width, the same trend is observed, but for decreasing values of $w$, as shown in figure \ref{S_fig:8in1_SiN}b. Again, the same is true for the thickness $h$, which is depicted in figure \ref{S_fig:8in1_SiN}.

\section{Absorber - supplementary results ($\tau_{th}$ and $R_P$)}
\begin{figure}
    \centering
    \includegraphics[width=1\linewidth, height=0.95\textheight]{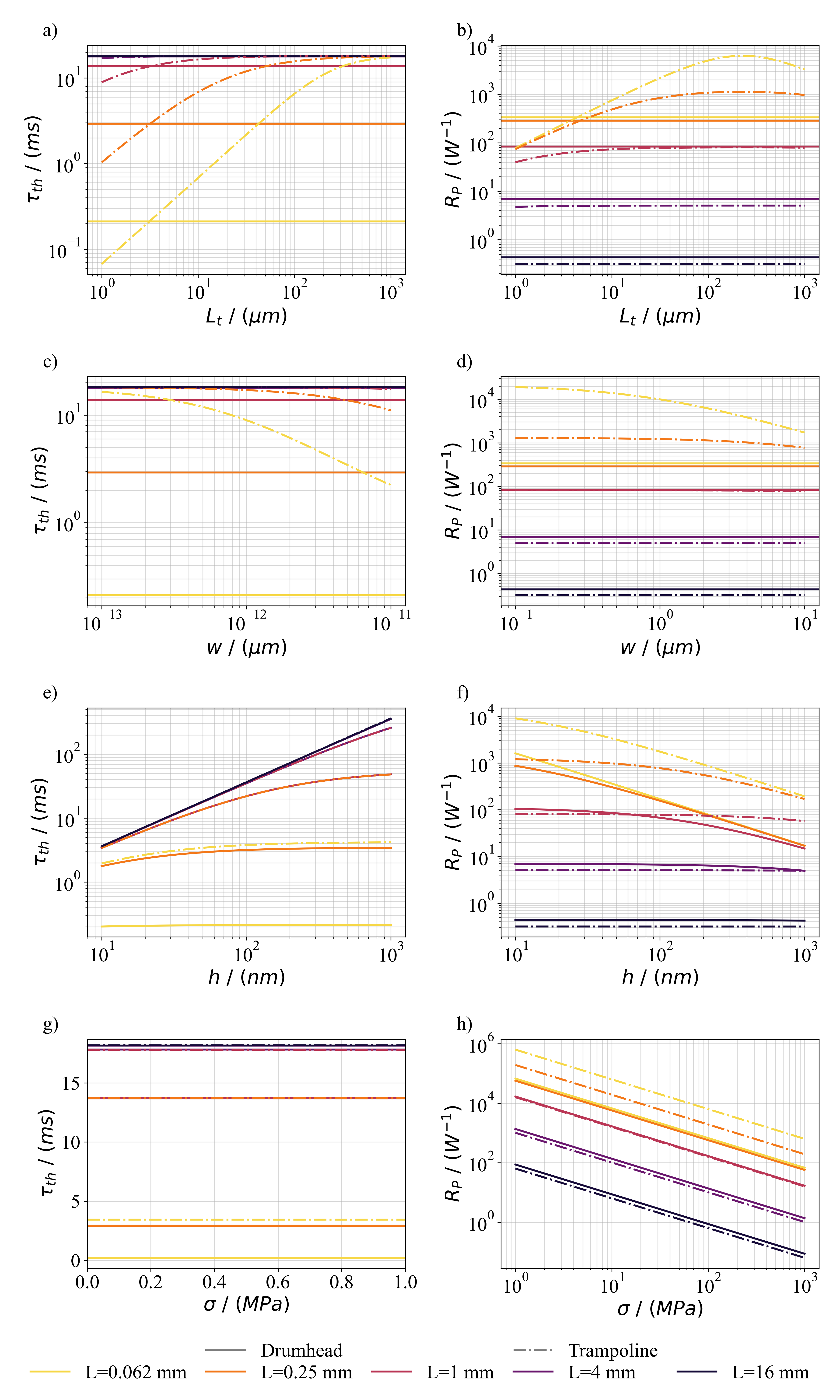}
    \caption{Results of $\tau_{th}$ and $R_P$ for varying a-b) tethers length $L_t$, c-d) tethers width $w$, e-f) thickness $h$, and g-h) tensile stress $\sigma$.}
    \label{S_fig:8in1_abs50}
\end{figure}

In Figure \ref{S_fig:8in1_abs50} we report the results for $\tau_{th}$ and $R_P$ in the case of a detector equipped with an ideal absorber, characterized by $\alpha_{abs} = \epsilon_{abs} = 50\%$. For all considered geometrical parameters, $L_t$, $w$, and $h$, the thermal time constant exhibits the same qualitative trend as in bare SiN. The key difference is that the values of $\tau_{th}$ are approximately one order of magnitude lower (see Figure \ref{S_fig:8in1_abs50}a, \ref{S_fig:8in1_abs50}c, and \ref{S_fig:8in1_abs50}e), owing to the enhanced emissivity introduced by the absorber. As discussed in the main text, the prestress $\sigma$ does not affect the thermal properties of the detectors, as evidenced by the horizontal lines in Figure \ref{S_fig:8in1_abs50}g.

Examining the results for the power responsivity $R_P$ in Figures \ref{S_fig:8in1_abs50}b, \ref{S_fig:8in1_abs50}d, \ref{S_fig:8in1_abs50}f, and \ref{S_fig:8in1_abs50}h, a clear trend emerges: across all variables, smaller structures exhibit higher $R_P$, with the smallest trampolines consistently representing the best overall performers. We attribute this to the enhanced thermal isolation of this particular design.

\section{Fabrication parameters for the works reported in Figure 12}
In Table \ref{tab:parameters}, we report the geometrical and stress details for the mechanical resonators we reported in Fig. 12.

\begin{table}[h!]
    \centering
    \begin{tabular}{l|c|c|c|c|c|}
        Detector & $L$ ($\mu$m) & $h$ (nm) & $L_t$ ($\mu$m) & $w$ ($\mu$m) & $\sigma$ (MPa) \\
        \hline \hline
        Membrane w/ absorber\cite{martini2025uncooled} & 1000 & 50 & - & - & 50 \\ \hline
        SiN Trampoline\cite{piller2022thermal} & 45 & 50 & 675 & 5 & 50 \\ \hline
        Trampoline w/ absorber\cite{zhang2025enhanced} & 300 & 90 & 636\footnotemark[1] & 130\footnotemark[1] & 70 \\ 
    \end{tabular}
    \footnotetext[1]{Values not reported. Extracted from image.}
    \caption{Geometrical and stress parameters of the compared devices.}
    \label{tab:parameters}
\end{table}

\bibliography{bib}